\documentclass[aps,prx,fixfloat,twocolumn]{revtex4-2}
\usepackage{graphicx,amsmath,amsfonts,bm, color}
\usepackage{wasysym}
\usepackage{tabularx}

\usepackage{xcolor,hyperref}
\hypersetup{
   colorlinks,
   linkcolor={blue!50!black},
   citecolor={blue!50!black},
   urlcolor={blue!80!black}
}

\usepackage{enumitem}

\renewcommand{\=}{\!=\!}
\newcommand{\1}{^{\mbox{\tiny (1)}}}

\newcommand{\tr}{\operatorname{tr}}

\newcommand{\dbar}{{\,\mathchar'26\mkern-12mu d}}
\DeclareMathAlphabet{\mathitbf}{OML}{cmm}{b}{it}

\newcommand{\xv}{\mathitbf x}

\newcommand{\calBold}[1]{\mbox{\boldmath${\cal #1}$}}
\def\onedot{$\mathsurround0pt\ldotp$}

\def\cddot{
  \mathbin{\vcenter{\baselineskip.67ex
    \hbox{\onedot}\hbox{\onedot}}}}

\begin{document}

\title{Elementary processes in dilatational plasticity of glasses}
\author{Avraham Moriel$^{1}$}
\author{David Richard$^{2}$}
\author{Edan Lerner$^{3}$}
\author{Eran Bouchbinder$^{1}$}
\affiliation{$^{1}$Chemical and Biological Physics Department, Weizmann Institute of Science, Rehovot 7610001, Israel\\
$^{2}$Universit\'{e} Grenoble Alpes, CNRS, LIPhy, 38000 Grenoble, France\\
$^{3}$Institute for Theoretical Physics, University of Amsterdam, Science Park 904, Amsterdam, Netherlands}

\begin{abstract}
Materials typically fail under complex stress states, essentially involving dilatational (volumetric) components that eventually lead to material decohesion/separation. It is therefore important to understand dilatational irreversible deformation --- i.e., dilatational plasticity --- en route to failure. In the context of glasses, much focus has been given to shear (volume-preserving) plasticity, both in terms of the stress states considered and the corresponding material response. Here, using a recently-developed methodology and extensive computer simulations, we shed basic light on the elementary processes mediating dilatational plasticity in glasses. We show that plastic instabilities, corresponding to singularities of the glass Hessian, generically feature both dilatational and shear irreversible strain components. The relative magnitude and statistics of the strain components depend both on the symmetry of the driving stress (e.g., shear vs.~hydrostatic tension) and on the cohesive (attractive) part of the interatomic interaction. We further show that the tensorial shear component of the plastic strain is generally non-planar and also extract the characteristic volume of plastic instabilities. Elucidating the fundamental properties of the elementary micro-mechanical building blocks of plasticity in glasses sets the stage for addressing larger-scale, collective phenomena in dilatational plasticity such as topological changes in the form of cavitation and ductile-to-brittle transitions. As a first step in this direction, we show that the elastic moduli markedly soften during dilatational plastic deformation approaching cavitation.
\end{abstract}

\maketitle

\section{Introduction}
\label{sec:intro}

Glassy materials are ubiquitous in the natural and technological world around us, and include various noncrystalline solids such as oxide glasses, glassy polymers, organic glasses and metallic glasses. These intrinsically disordered materials possess notable properties and hence find an enormous range of engineering applications~\cite{meyers2008mechanical,ashby2018materials,suryanarayana2017bulk,wang2004bulk,wondraczek2022advancing}.
Processing glassy materials~\cite{schroers2010processing} and more so their performance, durability and structural integrity in various applications require deep and fundamental understanding of their mechanical deformation and failure modes. Failure typically involves complex stress states, essentially involving dilatational (volumetric) components that eventually lead to material decohesion/separation. This is evident from extensive experimental observations regarding cavitation in glasses under a wide variety of failure conditions (e.g.,~\cite{wang2007nanoscale,jiang2008energy,jiang2014cryogenic,bouchaud2008fracture,shao2014direct,narasimhan2015fracture,singh2016cavitation,an2016toughness,shen2021observation,richard2023bridging}).

Despite its crucial importance, our current understanding of irreversible (plastic) dilatational deformation of glasses lags far behind the corresponding understanding of shear (deviatoric, volume-preserving) plasticity. Indeed, a lot of attention and research effort have been devoted to studying the fundamental micro-mechanics and statistical-mechanical properties of shear plasticity --- especially in the statistical physics community (e.g.,~\cite{Malandro_Lacks,maloney2004subextensive,barriers_lacks_maloney,lemaitre2006_avalanches,tanguy2006plastic,lemaitre_strain_rate_2009,athermal_elasticity_2009,tsamados2009local,yieldingrapid_pre_2010,steady_states_with_jacques,falk_review,rodney2011modeling,salerno_robbins,hufnagel2016deformation,liu2016driving,ozawa2018random,richard2020predicting,richard2021brittle,kapteijns2021unified,gonzalez2023variability}), with a few notable recent counterexamples (e.g.,~\cite{an2011atomistic,murali2011atomic,guan2013cavitation,chaudhuri2016structural,paul2020cavity,shimada2022gas,dattani2022universal,dattani2023cavitation,dattani2023athermal}). The meaning of ``shear plasticity'' here is two-fold; first, it refers to shear-driven plasticity, i.e., to studying plasticity under {\em driving conditions} (stress states) in which the shear/deviatoric component strongly dominates the dilatational/volumetric one. Second, it refers to the irreversible {\em material response}, which in previous works heavily focused on plastic shear strains mediated by the so-called Shear Transformation Zones (STZs)~\cite{Argon1979,kobayashi1980computer,maeda1981atomistic,falk_langer_stz}.

The pressing need to understand dilatational plasticity processes has already been recognized in several materials research communities, mainly in the context of metallic glasses~\cite{wang2007nanoscale,jiang2008energy,jiang2014cryogenic,bouchaud2008fracture,shao2014direct,narasimhan2015fracture,singh2016cavitation,an2016toughness,shen2021observation,richard2023bridging,wang2016direct,shi2014intrinsic}, leading to various experimental (e.g.,~\cite{jiang2008energy,jiang2014cryogenic}), simulational (e.g.,~\cite{shi2014intrinsic,wang2016direct}) and modeling (e.g.,~\cite{sun2010intrinsic,shi2014intrinsic}) efforts. It has also given rise to the conjecture that at the fundamental level there exist in addition to STZs (the carriers of shear plasticity) also Tension-Transformation-Zones (TTZs)~\cite{jiang2008energy,huang2014ductile,jiang2014cryogenic}, whose activation leads to atomic-scale quasi-cleavage~\cite{chen2011failure}. Yet, these important efforts have not been focused on the basic micro-mechanics, geometry and statistical-mechanics of dilatational plasticity.

\begin{figure}[h]
  \centering
  \includegraphics[width=0.43\textwidth]{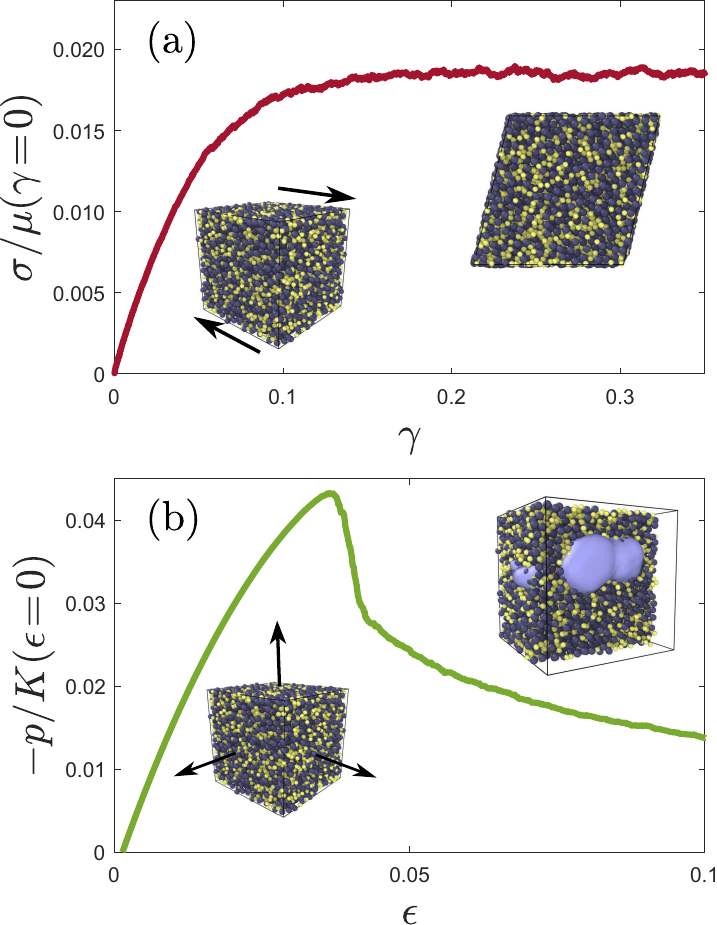}
\caption{(a) A shear stress-strain $\sigma(\gamma)$ of a 3D computer glass (of $N\!=\!10K$ particles, averaged over 108 realizations) deformed under simple shear. $\sigma$ is normalized by the initial shear modulus, $\mu(\gamma\!=\!0)$. The left inset shows the binary-mixture glass (blue and yellow particles correspond to the two species~\cite{SM}) prior to deformation and the arrows illustrate the subsequent application of simple shear. The right inset shows a side view of the same glass realization in the steady-state flow regime (here $\gamma\!\simeq\!0.25$). (b) The corresponding dilatational stress-strain $-p(\epsilon)$ for the same glass ensemble ($p$ is the pressure) deformed under hydrostatic tension. $-p$ is normalized by the initial bulk modulus, $K(\epsilon\!=\!0)$. The left inset shows the very same undeformed glass realization as in panel (a), but the arrows highlight the subsequent application of hydrostatic tension (pure dilation). The right inset shows a large-scale cavity (light blue region) inside the glass right after the observed abrupt stress drop. See text for a discussion of the results.}
  \label{fig:shear_vs_dilation}
\end{figure}

Our goal in this work is to study the fundamental micro-mechanics, geometric and statistical-mechanical properties of the elementary processes that mediate dilatational plasticity in glasses. In terms of driving forces, we invoke the hydrostatic tension test and compare it to its well-studied simple shear counterpart, both shown in Fig.~\ref{fig:shear_vs_dilation}. The hydrostatic tension test applied to computer glass samples physically represents a mesoscopic portion of a macroscopic glass, on which the surrounding material exerts predominantly hydrostatic forces. In Fig.~\ref{fig:shear_vs_dilation}a, an ensemble-averaged shear stress-strain curve under simple shear is presented, obtained through 3D computer simulations using athermal quasi-static (AQS) deformation~\cite{Malandro_Lacks,maloney2004subextensive,barriers_lacks_maloney,lemaitre2006_avalanches}. The resulting curve features a small-strain linear regime and a smooth, monotonic transition to a steady-state flow plateau upon shear plastic yielding. The curve does not feature a stress overshoot (and the accompanying stress drop), a situation that is typical for poorly annealed (i.e., rapidly quenched) glasses, which indeed corresponds to the employed glass ensemble~\cite{SM}.

In Fig.~\ref{fig:shear_vs_dilation}b, an ensemble-averaged dilatational stress-strain curve under hydrostatic tension (pure dilation) is presented, obtained using the very same ensemble of computer glasses as in panel (a). The resulting curve features a small-strain linear regime, followed by an abrupt stress drop and continuous strain softening. It qualitatively differs from its simple shear counterpart in Fig.~\ref{fig:shear_vs_dilation}a, despite performing the deformation simulations on the very same ensemble of glass realizations. One manifestation of this qualitative difference is the emergence of a cavitation instability in the hydrostatic tension test, as illustrated in the inset (see figure caption for additional details). A major goal of this work is to study the elementary irreversible processes that contribute to the observed differences.

Our methodology, involving computer glasses of $N$ particles and initial volume $V_0$, is discussed in Appendix~\ref{sec:methods}, where additional technical details are provided in~\cite{SM}. As material cohesion is essential for understanding dilatational plasticity and failure, our computer simulations employ a class of recently introduced potentials, which feature both repulsion and cohesion/attraction, where the strength of the attractive part is continuously adjustable through a dimensionless parameter $r_{\rm c}$~\cite{karmakar2011effect,sticky_spheres_part_1,sticky_spheres_part2,SM}. The smaller $r_{\rm c}$ is, the stronger the cohesive/attractive part of the interaction, as demonstrated in the inset of Fig.~\ref{fig:dil_dev_distributions}a. It was recently shown that reducing $r_{\rm c}$ can lead to a ductile-to-brittle transition~\cite{richard2021brittle}.

Simple shear AQS deformation in a given direction is controlled by a shear strain $\gamma$, where a representative simple shear stress-strain curve $\sigma(\gamma)$ is presented in Fig.~\ref{fig:shear_vs_dilation}a. Hydrostatic tension (pure dilation) is controlled by a dilatational (volumetric) strain $\epsilon$, representative hydrostatic tension stress-strain curve $-p(\epsilon)$ is presented in Fig.~\ref{fig:shear_vs_dilation}b ($p$ is the hydrostatic pressure). Unstable plastic eigenmodes ${\bm u}({\bm r})$ (where ${\bm r}$ is a position vector relative to the center of the eigenmode~\cite{SM}), corresponding to a zero crossing of an eigenvalue of the Hessian ${\calBold M}$, are identified during AQS deformation, controlled either by $\gamma$ or $\epsilon$ (see Appendix~\ref{sec:methods} and~\cite{SM}). ${\bm u}({\bm r})$ features a highly-nonlinear, disordered core of linear size $a$ (i.e., of volume ${\cal V}\!\propto\!a^3$) and a power-law decay $|{\bm r}|^{-2}$ in the far field (e.g.,~\cite{lemaitre2006_avalanches}), $|{\bm r}|\!\gg\!a$, associated with a linear elastic continuum behavior.

The irreversible deformation inside the nonlinear core is quantified through an eigenstrain tensor $\bm{\mathcal{E}}^*$ in the framework of Eshelby's inclusions formalism~\cite{eshelby1957determination,eshelby1959elastic}. A recently developed method~\cite{moriel2020extracting}, based on a class of contour integrals evaluated in the far field $|{\bm r}|\!\gg\!a$, allows to extract ${\cal V}\,\bm{\mathcal{E}}^*$. Our primary goal is to study the properties of $\bm{\mathcal{E}}^*$ and ${\cal V}$ (or alternatively $a$) as a function of the symmetry and magnitude of the driving force, quantified by $\gamma$ and $\epsilon$, and as a function of the strength of the cohesive/attractive part of the interatomic interaction, quantified by $r_{\rm c}$.

\section{Plastic eigenstrain triaxility}
\label{sec:triaxiality}

The eigenstrain tensor can be additively decomposed into its dilatational (volumetric) part, $\bm{\mathcal{E}}^*_{\mbox{\scriptsize dil}}$, and deviatoric (shear, volume-preserving) part, $\bm{\mathcal{E}}^*_{\mbox{\scriptsize dev}}$, $\bm{\mathcal{E}}^*\=\bm{\mathcal{E}}^*_{\mbox{\scriptsize dil}}+\bm{\mathcal{E}}^*_{\mbox{\scriptsize dev}}$. In 3D, $\bm{\mathcal{E}}^*$ is characterized by 3 independent amplitudes, one characterizing the dilatational (isotropic) part, $\bm{\mathcal{E}}^*_{\mbox{\scriptsize dil}}\=\epsilon_{\mbox{\scriptsize dil}}^*\,\bm{\mathcal{I}}$ and two characterizing the deviatoric part, which features $\tr(\bm{\mathcal{E}}^*_{\mbox{\scriptsize dev}})\=0$. We denote by $\epsilon_{\mbox{\scriptsize dev,1}}^*$ the largest (in absolute value) of the three eigenvalues of $\bm{\mathcal{E}}^*_{\mbox{\scriptsize dev}}$ and by $\epsilon_{\mbox{\scriptsize dev,2}}^*$ the second largest, which has a different sign compared to $\epsilon_{\mbox{\scriptsize dev,1}}^*$ (the third eigenvalue is not independent, but is rather determined through $\tr(\bm{\mathcal{E}}^*_{\mbox{\scriptsize dev}})\=0$).

The 3 independent amplitudes that characterize $\bm{\mathcal{E}}^*$ provide basic information about the geometry of plastic rearrangements/instabilities in glasses. Of particular interest in the present context is the amplitude of the dilatational eigenstrain component $\epsilon_{\mbox{\scriptsize dil}}^*$, and more specifically its relative magnitude compared to the deviatoric component. Consequently, we aim at constructing a ratio of the two components in order to quantify their relative magnitude. This goal naturally fits the contour integrals method~\cite{moriel2020extracting} that --- as noted above --- allows to extract only the product ${\cal V}\,\bm{\mathcal{E}}^*$, but not ${\cal V}$ and $\bm{\mathcal{E}}^*$ individually. In fact, as ${\bm u}({\bm r})$ is a normalized mode (by construction), considering absolute eigenstrain values is not immediate. Yet, as will be shown later, the fully nonlinear ${\bm u}({\bm r})$ does contain information that allows to estimate the core volume ${\cal V}$.

In order to construct a ratio that quantifies the relative magnitude of the dilatational and deviatoric components of the eigenstrain tensor $\bm{\mathcal{E}}^*$, we draw analogy with the well-known and widely-used stress triaxiality measure~\cite{davis1959stress,murakami2012continuum} --- constructed at the macroscopic scale for a similar goal, but with respect to the stress tensor --- and define
\begin{equation}
R_{\rm t} \equiv \frac{\epsilon_{\mbox{\scriptsize dil}}^*}{\sqrt{3J^*_2}} \ ,
\label{eq:triaxiality}
\end{equation}
where $J^*_2\!\equiv\!\tfrac{1}{2}\bm{\mathcal{E}}^*_{\mbox{\scriptsize dev}}\!\cddot\!\bm{\mathcal{E}}^*_{\mbox{\scriptsize dev}}$. We term the ratio $R_{\rm t}$ in Eq.~\eqref{eq:triaxiality} the plastic eigenstrain triaxiality. Note that $R_{\rm t}$ is a signed quantity, as will be further discussed below.
\begin{figure}[h]
  \centering
  \includegraphics[width=0.48\textwidth]{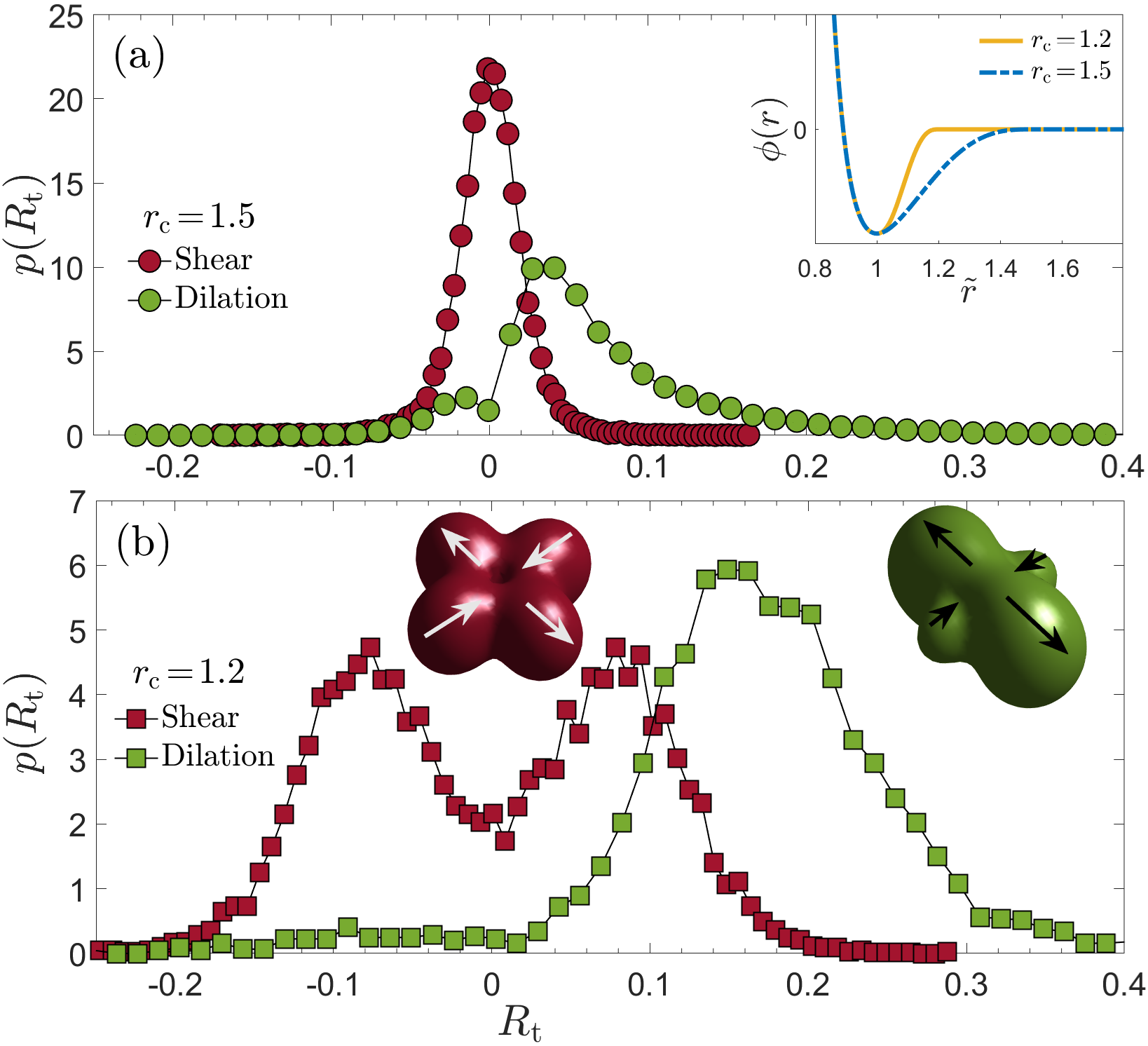}
  \caption{(a) The probability distribution $p(R_{\rm t})$ of the plastic eigenstrain triaxiliaty $R_{\rm t}$ defined in Eq.~\eqref{eq:triaxiality} for $r_{\rm c}\!=\!1.5$ glasses, under both simple shear and hydrostatic tension (see legend). The inset presents the interatomic pair interaction $\phi(r)$ vs.~$\tilde{r}$, where $r$ is the scaled distance between interacting particles and $\tilde{r}$ is further normalized such that the minimum of each curve occurs at $\tilde{r}\!=\!1$~\cite{sticky_spheres_part_1,SM}. Curves for the two values of $r_{\rm c}$ employed in this work (see legend) are shown. (b) The same as panel (a), but for $r_{\rm c}\!=\!1.2$. See text for an extensive discussion of the results. The visual insets present iso-surfaces of the magnitude of plastic modes with two values of $R_{\rm t}$. The left one (red), corresponds to $R_{\rm t}\!=\!0$ (i.e., $\epsilon_{\mbox{\scriptsize dil}}^*\!=\!0$) and a planar deviatoric eigenstrain tensor with $J^*_2\!=\!1$. The effective dimensionality of the deviatoric eigenstrain tensor (quantifying its degree of planarity) is discussed in relation to Fig.~\ref{fig:planarity}. The arrows (white) indicate the direction of the displacement (with length that is consistent with its magnitude). The right visual inset (green) corresponds to $\epsilon_{\mbox{\scriptsize dil}}^*\!=\!0.5$ and again a planar deviatoric eigenstrain tensor with $J^*_2\!=\!1$, resulting in $R_{\rm t}\!\simeq\!0.29$. The arrows (black) indicate the direction of the displacement (with length that is consistent with its magnitude).}
  \label{fig:dil_dev_distributions}
\end{figure}

In Fig.~\ref{fig:dil_dev_distributions}, the probability distribution $p(R_{\rm t})$ is plotted for both the simple shear and hydrostatic tension tests, and two values of $r_{\rm c}$. $p(R_{\rm t})$ was obtained by calculating $R_{\rm t}$ of Eq.~\eqref{eq:triaxiality} for 200 independent glass samples (made of $N\=128$K particles each) per $r_{\rm c}$ value, where for each sample up to the first 50 plastic instabilities/events were detected~\cite{SM}. Under dilation, we focused on strains below the cavitation strain (corresponding to the peak stress under hydrostatic tension, cf.~Fig.~\ref{fig:shear_vs_dilation}b), implying that not all of the detected plastic events were analyzed. Overall, the presented distributions were generated using between 3000 and 10000 events each. Statistical convergence and possible effects of the magnitude of the applied strain are discussed in~\cite{SM}, also for other observables to be considered below.

In Fig.~\ref{fig:dil_dev_distributions}a, we show $p(R_{\rm t})$ for $r_{\rm c}\=1.5$, which corresponds to the canonical Lennard-Jones interatomic potential (see inset). It is observed that under simple shear (red circles) $p(R_{\rm t})$ is symmetric and narrowly peaked around $R_{\rm t}\=0$ (i.e., a vanishing dilatational eigenstrain, $\epsilon_{\mbox{\scriptsize dil}}^*\=0$). Note that $\epsilon_{\mbox{\scriptsize dil}}^*\!>\!0$ corresponds to isotropic core expansion/dilation and $\epsilon_{\mbox{\scriptsize dil}}^*\!<\!0$ to isotropic core contraction~\cite{SM}. It is observed that finite values of $\epsilon_{\mbox{\scriptsize dil}}^*$ exist, though $R_{\rm t}$ is small, indicating that the deviatoric eigenstrain component is significantly larger than the dilatational one. Under hydrostatic tension (green circles), for the very same ensemble of glass realizations, the symmetry between core expansion and contraction, $R_{\rm t}\!\to\!-R_{\rm t}$, is broken, and $p(R_{\rm t})$ is biased toward positive $R_{\rm t}$ values. $p(R_{\rm t})$ is peaked at relatively small values, again indicating a dominance of the deviatoric component over the dilatational one, even when the global driving force is dilatational (hydrostatic tension).

The results presented in Fig.~\ref{fig:dil_dev_distributions}a demonstrate that plastic instabilities in glasses generically feature both dilatational and shear components. A corollary of this finding is that at least part of dilatational plasticity is carried by the same micro-mechanical objects that carry shear plasticity. Moreover, the relative magnitude of the dilatational and deviatoric components depends on the stress state, i.e., the microscopic plastic strain is not an intrinsic/geometric property of plastic instabilities (for a given glass composition and interatomic interaction). Finally, hydrostatic tension gives rise to larger dilatational plastic strains, though for the canonical Lennard-Jones potential the deviatoric plastic strain is dominant.

We then repeated the calculations leading to Fig.~\ref{fig:dil_dev_distributions}a for an ensemble of glasses generated with $r_{\rm c}\=1.2$. This value of $r_{\rm c}$ corresponds to an interatomic potential featuring a stronger cohesive/attractive part, see inset of Fig.~\ref{fig:dil_dev_distributions}a. In the presence of stronger cohesion, one expects that in order to trigger an irreversible rearrangement, particles at the core of a plastic mode (saddle configuration) would feature a larger separation, i.e., $p(R_{\rm t})$ is expected to feature larger values. The results in Fig.~\ref{fig:dil_dev_distributions}b support this expectation. 

It is observed that under simple shear (red squares) $p(R_{\rm t})$ is still symmetric around $R_{\rm t}\=0$, but becomes bi-modal with peaks at $R_{\rm t}$ values significantly larger than the typical $R_{\rm t}$ values for $r_{\rm c}\=1.5$ (cf.~the red circles in Fig.~\ref{fig:dil_dev_distributions}a). Moreover, this physical effect significantly intensifies under hydrostatic tension, as demonstrated by $p(R_{\rm t})$ in Fig.~\ref{fig:dil_dev_distributions}b (green squares). It is observed that under these conditions, $r_{\rm c}\=1.2$ glasses feature almost only positive $R_{\rm t}$ values, and $p(R_{\rm t})$ is broad, where plastic events with a sizable dilatational component exist with a non-negligible probability. The geometry of plastic cores with different $R_{\rm t}$ values is illustrated in the visual insets of Fig.~\ref{fig:dil_dev_distributions}b, see figure caption for details. 

Overall, the results of Fig.~\ref{fig:dil_dev_distributions} show that a dilatational plastic strain generically exists in plastic instabilities in glasses, along with a deviatoric plastic strain, and that their relative magnitude depends on both the driving conditions and the cohesive/attractive part of interatomic interactions. Moreover, our findings show that at least part of dilatational plasticity in glasses is carried by the same micro-mechanical objects that carry shear plasticity, and consequently that these provide a coupling mechanism between the two.

The generic existence of a dilatational plastic strain in plastic events in glasses has also been demonstrated previously in 2D Lennard-Jones glasses studied under uniaxial tension/compression~\cite{ashwin2013yield} and under simple shear~\cite{nicolas2018orientation}, and in 3D computer models of amorphous silicon studied under simple shear~\cite{albaret2016mapping}. Amorphous silicon has been modelled in~\cite{albaret2016mapping} using either the standard/original Stillinger-Weber (SW) potential~\cite{stillinger1985computer} or a modified SW (termed SWM therein) potential~\cite{holland1998ideal,holland1999cracks}. The latter potential, which features a three-body term twice as large as the original SW potential, has been developed in relation to the fracture of crystalline silicon, where it was found that the SW potential leads to a ductile behavior, while the SWM potential to a brittle one~\cite{holland1998ideal,holland1999cracks}. This is reminiscent of the ductile-to-brittle transition induced by reducing $r_{\rm c}$, as discussed in~\cite{richard2021brittle}. The ductile-to-brittle transition in both cases is also accompanied by a reduction in Poisson's ratio (see Table I in~\cite{albaret2016mapping} for the SW and SWM values, and Table S2 in~\cite{SM} for the different $r_{\rm c}$ values).

Interestingly, it was reported in~\cite{albaret2016mapping} (cf.~Fig.~7 therein) that the relative magnitude of the dilatational and deviatoric components of plastic events under simple shear (termed ``shear-tension coupling'') increases from SW to SWM. This is precisely the trend observed in Fig.~\ref{fig:dil_dev_distributions} (red symbols in the two panels) with decreasing $r_{\rm c}$. Consequently, while ``ductility'' and ``brittleness'' are clearly collective phenomena that are affected by both the history dependence of a glass (i.e., its initial state of disorder emerging from the quench self-organization, being in itself affected by the interatomic potential) and by spatiotemporal interactions of an extensive number of plastic events during deformation~\cite{richard2021brittle,richard2023bridging}, they might also have some signature in the geometry of individual plastic events.

\section{Planarity of the deviatoric eigenstrain tensor}
\label{sec:planarity}

The plastic eigenstrain triaxility $R_{\rm t}$, studied above, constructs out of the 3 independent amplitudes that characterize the eigenstrain tensor $\bm{\mathcal{E}}^*$ a measure of the relative magnitude of the dilatational and deviatoric eigenstrain components. Next, we consider another geometric property of the core of plastic instabilities. As the dilatational component is isotropic, i.e., it features 3D spherical geometry, we focus our attention on the geometry of the deviatoric eigenstrain tensor $\bm{\mathcal{E}}^*_{\mbox{\scriptsize dev}}$, which is characterized by the 2 independent amplitudes $\epsilon_{\mbox{\scriptsize dev,1}}^*$ and $\epsilon_{\mbox{\scriptsize dev,2}}^*$, defined above.

A lot of previous insight into glass plasticity has been gained using computer simulations in 2D, which --- as highlighted above --- focused mostly on shear plasticity. In 2D, the deviatoric eigenstrain tensor is characterized by a single independent amplitude. There were some preliminary indications in the literature that plastic instabilities in 3D feature such a planar deviatoric eigenstrain tensor as well (e.g.,~\cite{carmel_pre_2013}). That is, in terms of the above-defined quantities, it was suggested that $\epsilon_{\mbox{\scriptsize dev,1}}^*\!\simeq\!-\epsilon_{\mbox{\scriptsize dev,2}}^*$ (such that the magnitude of the third eigenstrain is much smaller than $|\epsilon_{\mbox{\scriptsize dev,1}}^*|$) also characterizes plastic instabilities in 3D. Yet, to the best of our knowledge, this issue has not been systematically investigated in the past.

\begin{figure}[h]
  \centering
  \includegraphics[width=0.48\textwidth]{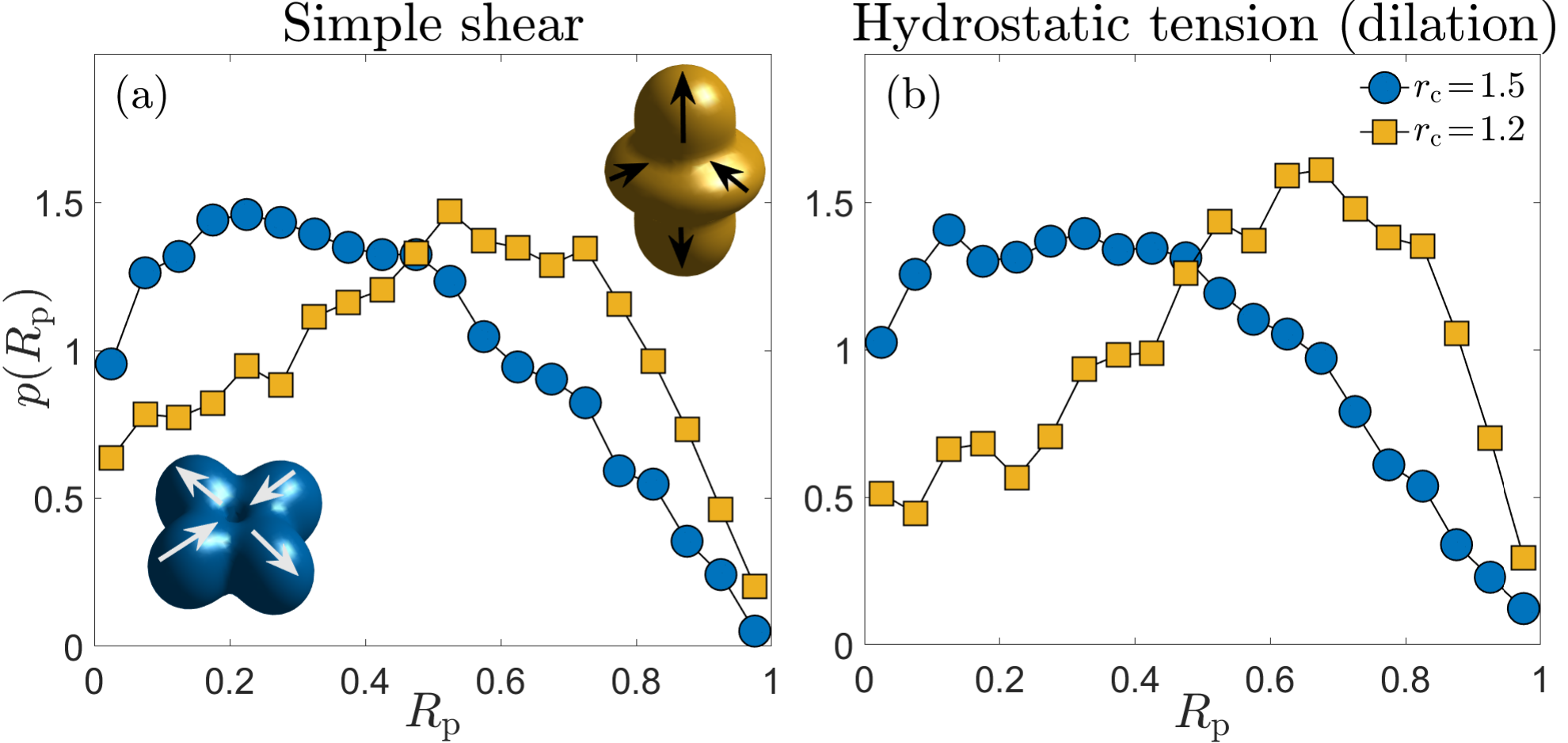}
  \caption{The probability distribution $p(R_{\rm p})$ of the planarity ratio $R_{\rm p}$ of the deviatoric eigenstrain tensor, cf.~Eq.~\eqref{eq:planarity}, in glasses under simple shear, for two values of $r_{\rm c}$ (see legend in panel (b)). The visual insets present iso-surfaces of the magnitude of the deviatoric part of plastic modes with two values of $R_{\rm p}$. The left one (blue), corresponds to $R_{\rm p}\!=\!0$ (i.e., the purely planar limit), and is identical to the left visual inset in Fig.~\ref{fig:dil_dev_distributions}b. The right visual inset (yellow), corresponds to $R_{\rm p}\!=\!1$ (i.e., the least planar). In both cases, the arrows indicate the direction of the displacement (with length that is consistent with its magnitude). (b) The same as panel (a), but under hydrostatic tension. See text for a discussion.}
  \label{fig:planarity}
\end{figure}

To address this basic issue, we define the planarity ratio of the deviatoric eigenstrain tensor $\bm{\mathcal{E}}^*_{\mbox{\scriptsize dev}}$ of plastic instabilities as
\begin{equation}
R_{\rm p}\!\equiv\!2\left(1+\frac{\epsilon_{\mbox{\scriptsize dev,2}}^*}{\epsilon_{\mbox{\scriptsize dev,1}}^*}\right) \ .
\label{eq:planarity}
\end{equation}
As explained above, in the purely planar limit one eigenvalue of $\bm{\mathcal{E}}^*_{\mbox{\scriptsize dev}}$ vanishes and we have $\epsilon_{\mbox{\scriptsize dev,1}}^*\=-\epsilon_{\mbox{\scriptsize dev,2}}^*$, i.e., $R_{\rm p}\=0$. The opposite limit, the least planar one, corresponds to $\epsilon_{\mbox{\scriptsize dev,2}}^*/\epsilon_{\mbox{\scriptsize dev,1}}^*\=-1/2$, i.e., $R_{\rm p}\=1$. These limiting cases are illustrated in the visual insets in Fig.~\ref{fig:planarity}a (see figure caption for additional details). In these terms, the suggestion that the deviatoric eigenstrain tensor remains planar in 3D corresponds to $p(R_{\rm p})\!\to\!\delta(R_{\rm p})$, i.e., to a probability distribution $p(R_{\rm p})$ that is strongly concentrated near the planar limit $R_{\rm p}\!\simeq\!0$, approaching a delta-function distribution. Our goal is to study $p(R_{\rm p})$ as a function of the external driving force (simple shear vs.~hydrostatic tension) and $r_{\rm c}$.

The analysis was performed on the same set of plastic instabilities/events discussed in relation to Fig.~\ref{fig:dil_dev_distributions}, and the results are presented in Fig.~\ref{fig:planarity}. In Fig.~\ref{fig:planarity}a, $p(R_{\rm p})$ under simple shear is plotted for two values of $r_{\rm c}$ (those previously used in Fig.~\ref{fig:dil_dev_distributions}, see legend in Fig.~\ref{fig:planarity}b). It is observed that for $r_{\rm c}\=1.5$, $p(R_{\rm p})$ features a broad peak around $R_{\rm p}\!\simeq\!0.3$, significantly deviating from $\delta(R_{\rm p})$. Moreover, for $r_{\rm c}\=1.2$ (corresponding to a stronger attractive part of the interatomic potential), $p(R_{\rm p})$ is also broad, but is peaked at significantly larger values of $R_{\rm p}$, away from the planar limit. The corresponding results under hydrostatic tension are presented in Fig.~\ref{fig:planarity}b, and are similar (note that $p(R_{\rm p})$ is more sharply peaked around $R_{\rm p}\!\simeq\!0.7$ compared to the corresponding result under simple shear, cf.~panel (a)). Overall, the results indicate that the geometry of the deviatoric eigenstrain tensor $\bm{\mathcal{E}}^*_{\mbox{\scriptsize dev}}$ is generally non-planar, i.e., that the dimensionality of $\bm{\mathcal{E}}^*_{\mbox{\scriptsize dev}}$ is the same as space dimensionality.

\section{The core size of plastic instabilities}
\label{sec:core_size}

As pointed out above, the contour integrals method, which employs the far field linear elastic properties of unstable plastic modes ${\bm u}({\bm r})$ to infer the core properties, does not allow to separately extract the core volume ${\cal V}$ and plastic eigenstrain tensor $\bm{\mathcal{E}}^*$. Yet, the core volume ${\cal V}$ --- or equivalently the linear core size $a$ (with ${\cal V}\!\propto\!a^3$) --- is an important glassy lengthscale. For example, it has been argued that ${\cal V}$ influences various physical properties of glasses~\cite{pan2009correlation,jiang2011shear,chen2016clarification}. Consequently, it is important to extract ${\cal V}\!\propto\!a^3$ itself. In fact, the particle-level (atomistic), fully nonlinear ${\bm u}({\bm r})$ can be used to estimate $a$.

The unstable plastic mode displacement field ${\bm u}({\bm r})$ is evaluated at each spatial position ${\bm r}_i$ occupied by a particle, hence it can be denoted as ${\bm u}_i\!\equiv\!{\bm u}({\bm r}_i)$ (which is the quantity we actually calculate to begin with). ${\bm u}_i$ follows a continuum linear elastic behavior for $|{\bm r}_i|\!\gg\!a$, yet it features significant nonlinearity and larger displacements over shorter distances. A widely used measure of spatial localization is the participation number (i.e., the participation ratio multiplied by $N$) $(\sum_i\!|{\bm u}_i|^2)^2/\sum_i\!|{\bm u}_i|^4$~\cite{bell1970atomic,bell1972dynamics}, which is ${\cal O}(1)$ in the extreme localization limit and ${\cal O}(N)$ in the spatially extended limit. Consequently, the participation number provides an estimate for the number of particles inside the core of ${\bm u}_i$, and we estimate the dimensionless core size as
\begin{equation}
\frac{a}{a_0} \equiv \left(\frac{1}{\sum_i\!|{\bm u}_i|^4}\right)^{1/3} \ .
\label{eq:core_size}
\end{equation}
Here, $a_0\!\equiv\!(V_0/N)^{1/3}$ is a characteristic interparticle distance, $i\=1\!-\!N$ runs over all particles and we used the fact that unstable plastic modes are normalized by construction, i.e., $\sum_i\!|{\bm u}_i|^2\=1$.
\begin{figure}[h]
  \centering
  \includegraphics[width=0.495\textwidth]{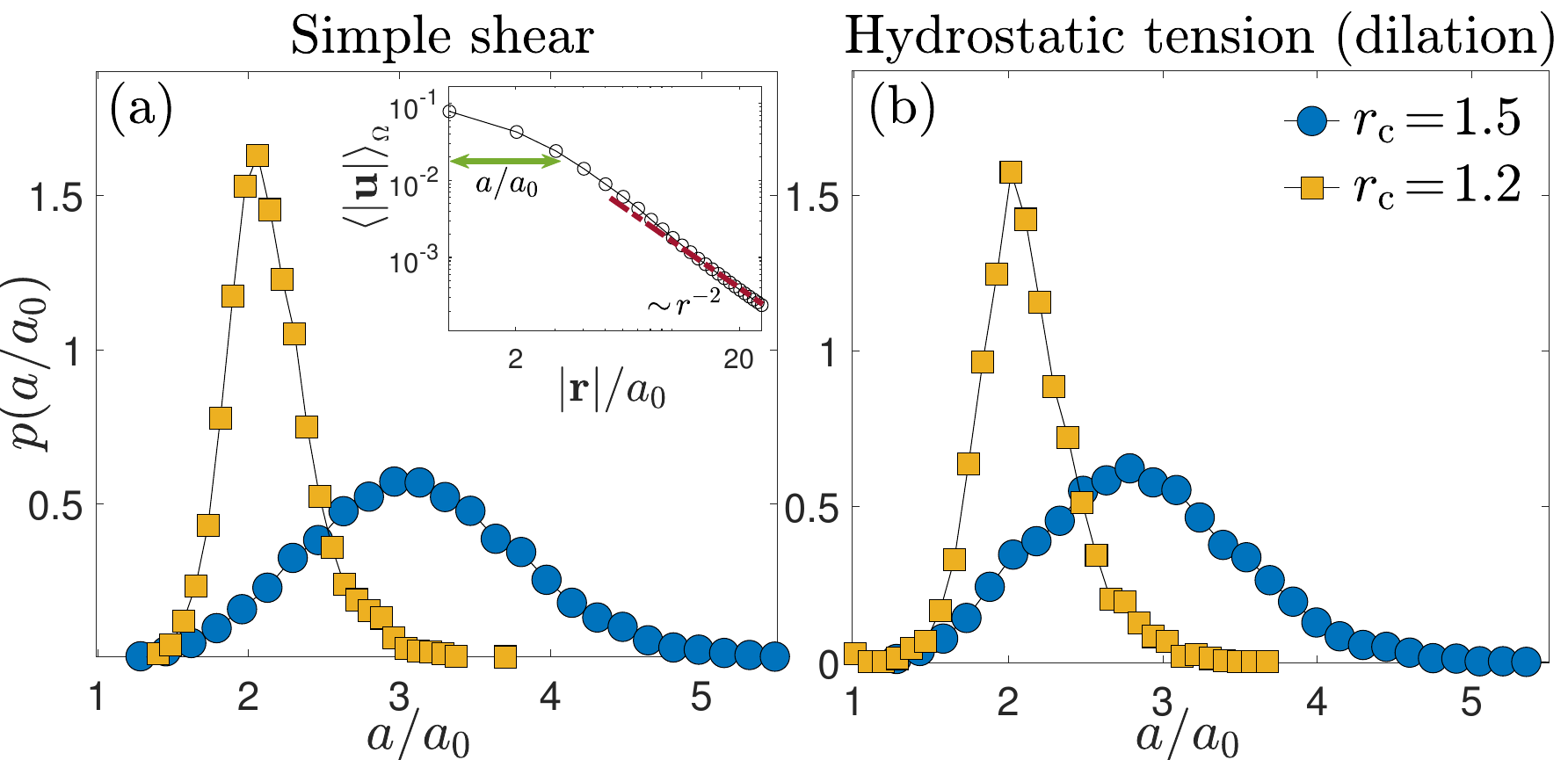}
  \caption{(a) The probability distribution $p(a/a_0)$ of the dimensionless linear size $a/a_0$ of the core of unstable plastic modes, as defined in Eq.~\eqref{eq:core_size}, for two values of $r_{\rm c}$ (see legend in panel (b)) under simple shear. (inset) The amplitude of an unstable plastic mode $\langle|{\bm u}|\rangle_{_\Omega}$ (the solid angle $\Omega$ average of the norm of ${\bm u}({\bm r})$) vs.~$|{\bm r}|/a_0$ in an $r_{\rm c}\!=\!1.5$ glass. The estimated core size $a/a_0$ is marked by the horizontal green double arrow and the linear elastic far-field power-law $\sim\!1/r^2$ is highlighted (red dashed-dotted line). (b) The same as panel (a), but under hydrostatic tension. See text for a discussion.}
  \label{fig:volume}
\end{figure}

The probability distribution $p(a/a_0)$ is presented in Fig.~\ref{fig:volume} for two values of $r_{\rm c}$, and under both simple shear and hydrostatic tension, for the plastic events previously analyzed in Figs.~\ref{fig:dil_dev_distributions}-\ref{fig:planarity}. In Fig.~\ref{fig:volume}a, we plot $p(a/a_0)$ under simple shear for both $r_{\rm c}\=1.5$ (light blue circles) and $r_{\rm c}\=1.2$ (yellow squares). It is observed that $p(a/a_0)$ is peaked at a few interparticle distances. Moreover, it shifts to larger values and becomes wider with increasing $r_{\rm c}$. Note that in the definition in Eq.~\eqref{eq:core_size} we do not account for the increase in the volume per particle in dilation (but rather use the fixed linear size $a_0$) because over the range of dilatational strains we consider, the implied changes are small.

In Fig.~\ref{fig:volume}b, we present the corresponding results under hydrostatic tension, which are quantitatively similar to the simple shear results presented in panel (a). The inset in panel (a) shows an example of the amplitude of an unstable plastic mode $\langle|{\bm u}|\rangle_{_\Omega}$ (the solid angle $\Omega$ average of the norm of ${\bm u}({\bm r})$) as a function of $|{\bm r}|/a_0$. The estimated core size $a/a_0$ is marked by the horizontal green double arrow and the linear elastic far-field power-law $\sim\!r^{-2}$ is highlighted (red dashed-dotted line). The results of Fig.~\ref{fig:volume} show that the core size of plastic instabilities is nearly independent of the symmetry of the loading, but does depend on the interatomic interaction potential. Specifically, it becomes more compact with increasing attractive forces, i.e., decreasing $r_{\rm c}$. This trend is also consistent with observed correlations between decreasing Poisson's ratio $\nu$ and the plastic core size (sometimes termed the STZ size/volume)~\cite{albaret2016mapping,sticky_spheres_part_1,pan2008experimental}, see the variation of $\nu$ with $r_{\rm c}$ in Table II of~\cite{SM}. 

The characteristic core size $a$ is similar to the core size $\xi_{\rm g}$ of quasilocalized, nonphononic modes in glasses~\cite{modes_prl_2016,ikeda_pnas,modes_prl_2018,LB_modes_2019,modes_prl_2020,QLE_jcp_2021}. Quasilocalized glassy modes, defined in the reference/undeformed glass, have been recently shown to follow a universal density of states $\sim\!\omega^4$ (distinct from Debye's density of states of low-frequency phonons, where $\omega$ is the vibrational frequency), and to feature a spatial structure similar to that of unstable plastic modes. In particular, quasilocalized modes are characterized by a nonlinear, disordered core of linear size $\xi_{\rm g}$ and a linear elastic power-law decay $|{\bm r}|^{-2}$ in the far field, $|{\bm r}|\!\gg\!\xi_{\rm g}$, exactly like unstable plastic modes. Moreover, the trend of variation of $a$ and $\xi_{\rm g}$ with varying $r_{\rm c}$ is similar (compare our results to Fig.~7i in~\cite{sticky_spheres_part_1}). Overall, our findings reinforce the suggestion that plastic instabilities are predominantly quasilocalized modes that are driven to a saddle-node bifurcation.

\section{The softening of elastic moduli under dilatational plasticity}
\label{sec:softening}

We provided above a basic quantitative characterization of the geometry and statistical-mechanical properties of the elementary micro-mechanical carriers of plasticity in glasses, for two end members of driving force symmetries (simple shear vs.~hydrostatic tension) and variable strength of cohesive/attractive interatomic interactions. In particular, the distributions of geometric properties (quantified by the ratios $R_{\rm t}$ and $R_{\rm p}$) and a characteristic length (core size $a$) of plastic instabilities have been calculated. Our focus was on dilatational plasticity and its comparison to shear plasticity, both in terms of the driving forces and the material response manifested in the localization length $a$ and the eigenstrain tensor $\bm{\mathcal{E}}^*$.

The collective spatiotemporal accumulation of these elementary plasticity processes, including their coupling and emerging spatial organization, gives rise to larger-scale plasticity in glasses. Much of these collective phenomena, in relation to dilatational plasticity in particular, remain to be explored and understood. We stress (again) that we focused on unstable plastic modes and not on the outcome of the instabilities, and also note that we do not claim to have exhaustively identified all elementary dilatational plasticity processes in glasses. For example, we have not discussed micro-cavitation, i.e., the process by which cavities on the particle scale are formed (to be distinguished from the larger-scale cavitation observed in Fig.~\ref{fig:shear_vs_dilation}b). These may be related to the micro-mechanical objects we identified (e.g., a subset of the outcomes of the identified plastic instabilities) or correspond to different objects. Yet, we would like to conclude this paper by demonstrating the type of new effects associated with collective dilatational plasticity.
\begin{figure}[h]
  \centering
  \includegraphics[width=0.48\textwidth]{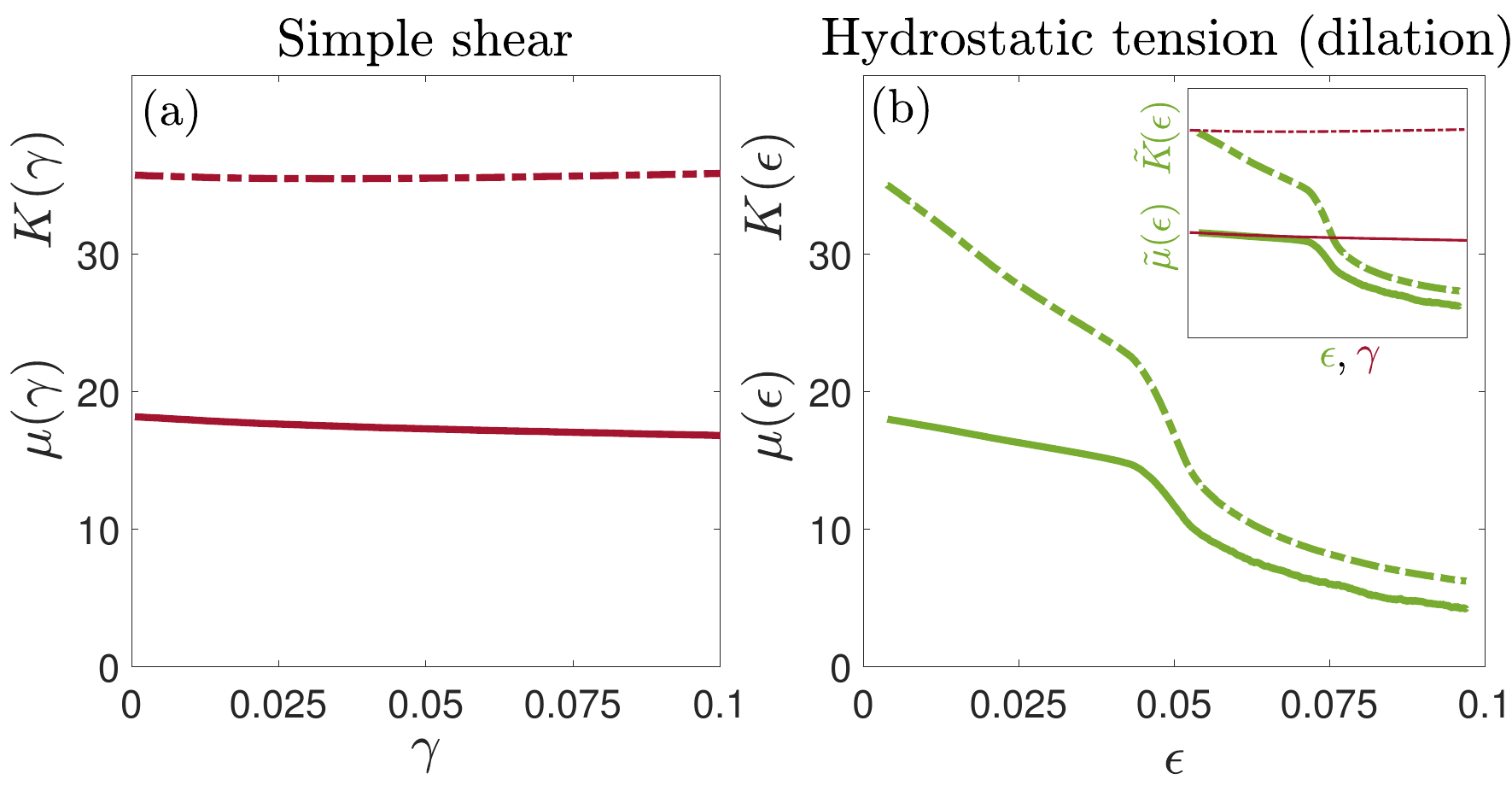}
  \caption{(a) The bulk $K(\gamma)$ (dashed-dotted line) and shear $\mu(\gamma)$ (solid line) moduli under simple shear for an ensemble of $r_{\rm c}\!=\!1.2$ glasses. The moduli, whose precise definition is given in~\cite{SM}, are reported in units of the interatomic interaction energy scale divided by the atomic volume $a_0^3$. (b) The same as panel (a), but under hydrostatic tension. (inset) The reduced moduli, $\tilde\mu(\epsilon)\!\equiv\!\mu(\epsilon)V(\epsilon)/V_0$ and $\tilde K(\epsilon)\!\equiv\!K(\epsilon)V(\epsilon)/V_0$, under hydrostatic tension (green labels and curve). The corresponding moduli under simple shear (red curves and $x$ label), whose values are identical to those presented in panel (a) since $V(\gamma)\!=\!V_0$, are superimposed for comparison. See text for a discussion of the results.}
  \label{fig:softening}
\end{figure}

We already demonstrated a qualitative difference between collective shear and dilatational elasto-plastic dynamics in Fig.~\ref{fig:shear_vs_dilation}, manifested at the level of stress-strain curves. Here, we provide another example, focusing on the strain evolution of elastic moduli, which can serve as global proxies for the structural changes experienced by a glass during elasto-plastic deformation. In Fig.~\ref{fig:softening}, we present the strain evolution of the ensemble-averaged bulk modulus $K$ and shear modulus $\mu$, under simple shear and hydrostatic tension. We stress that $K$ and $\mu$, which characterize minima of the potential energy (see precise definitions in~\cite{SM}), should not be confused with the tangent moduli, which characterize derivatives of the global, fluctuations averaged stress-strain curves. The two loading protocols are applied to the very same glass ensemble, composed of 200 samples generated with $N\=128$K and $r_{\rm c}\=1.2$ (e.g., compared to $N\=10$K and $r_{\rm c}\=1.5$ in Fig.~\ref{fig:shear_vs_dilation}), resulting in strains approaching the steady-state regime in the simple shear case and going beyond the stress drop in the hydrostatic tension case (cf.~Fig.~\ref{fig:shear_vs_dilation}).

It is observed that under simple shear (Fig.~\ref{fig:softening}a), both elastic moduli are largely unaffected by plastic deformation. In particular, the bulk modulus $K(\gamma)$ is essentially independent of $\gamma$ and $\mu(\gamma)$ softens by $\sim\!7\%$ from the initial (as cast, undeformed) glass to the steady shear flow. The corresponding results under hydrostatic tension are presented in Fig.~\ref{fig:softening}b. First, it is observed that both moduli experience a large drop associated with large-scale cavitation (see Fig.~\ref{fig:shear_vs_dilation}b), here around a cavitation strain of $\epsilon_{\rm c}\!\simeq\!0.045$. Second, both moduli significantly soften prior to $\epsilon_{\rm c}$, where the softening is more pronounced for the bulk modulus.

We note that the expressions for the elastic moduli include an overall factor $1/V$~\cite{SM}, where $V$ is the current volume. Under simple shear, which is volume preserving, we have $V(\gamma)\=V_0$. However, under hydrostatic tension, $V(\epsilon)\!\ge\!V_0$ is an increasing function of the dilatational strain $\epsilon$. It would be therefore interesting to disentangle the contribution of $1/V(\epsilon)$ to the observed softening under hydrostatic tension from other contributions by defining the reduced moduli, $\tilde\mu(\epsilon)\!\equiv\!\mu(\epsilon)V(\epsilon)/V_0$ and $\tilde K(\epsilon)\!\equiv\!K(\epsilon)V(\epsilon)/V_0$. The results are presented in the inset of Fig.~\ref{fig:softening}b. Interestingly, by superimposing the corresponding moduli under simple shear, which are identical to those presented in Fig.~\ref{fig:softening}a, it is observed that the reduced shear modulus almost coincides under both loading symmetries, prior to cavitation (in the hydrostatic tension case).
On the other hand, the reduced bulk modulus under hydrostatic tension still significantly deviates from its simple shear counterpart,
indicating the existence of intrinsic softening processes on top of the varying volume. This observation demonstrates qualitative differences between the shear and bulk moduli.

The softening of the elastic moduli emerges from the accumulation of dilatational plastic deformation and possibly micro-cavitation in a way that is not yet understood. Understanding these softening processes will also shed light on the qualitative differences between the shear and bulk moduli. More generally, understanding the spatiotemporal dynamics of collective dilatational plasticity upon approaching large-scale cavitation is a challenge for future work.

\section{Brief summary and outlook}
\label{sec:summary}

In this work, we studied elementary processes in glass plasticity, with a focus on dilatational plasticity and its comparison to its well-studied shear plasticity counterpart. We have extracted, using large-scale AQS computer simulations, the basic micro-mechanics, geometry and statistical-mechanical properties of unstable plastic modes (zero crossing of an eigenvalue of the glass Hessian) as a function of both the symmetry of the applied driving forces (simple shear vs.~hydrostatic tension) and the strength of the cohesive/attractive part of the interatomic interaction. These modes feature an effective Eshelby eigenstrain tensor over a spatial scale that defines their plastic core.

In particular, we computed 3 probability distribution functions of the following quantities: (i) the plastic eigenstrain triaxility $R_{\rm t}$ (a dimensionless measure of the relative magnitude of the dilatational and deviatoric parts of the eigenstrain tensor), (ii) the planarity ratio $R_{\rm p}$ (of the deviatoric part of the eigenstrain tensor) and (iii) the linear core size $a$. We found that $R_{\rm t}$ strongly depends on the symmetry of the applied driving forces and on the strength of the cohesive/attractive part of the interatomic interaction (Fig.~\ref{fig:dil_dev_distributions}). We also found that the deviatoric part of the eigenstrain tensor is generally non-planar, and that the statistical deviation from planarity is larger for more cohesive/attractive interatomic interactions (Fig.~\ref{fig:planarity}). Finally, we found that the statistics of $a$ are almost independent of the symmetry of the applied driving forces, but dependent on the strength of the cohesive/attractive part of the interatomic interaction (Fig.~\ref{fig:volume}). The latter results provide additional support for the intrinsic relations between unstable plastic modes and nonphononic modes in glasses~\cite{Zylberg2017,richard2020predicting,QLE_jcp_2021}.

At larger scales, involving the collective, spatially-coupled dynamics of many elementary plasticity processes, increasing hydrostatic tension leads to a cavitation instability upon which internal surfaces are formed. Cavity formation is accompanied by a large tension drop, but not an entire loss of load bearing capacity, which persists to larger dilatational strains (Fig.~\ref{fig:shear_vs_dilation}b). Interestingly, large-scale cavitation is preceded by a significant softening of the elastic moduli (Fig.~\ref{fig:softening}).

These results open the way for various research directions in dilatational plasticity, which can be roughly classified into two categories. First, additional insight into elementary plasticity processes should be gained. It remains to be understood whether the initial glassy state of disorder, e.g., as controlled by the quench rate through the glass transition temperature, affects the core properties of unstable plastic modes or just their occurrence probability~\cite{richard2021brittle}.

In this work, we studied unstable plastic modes, which are well-defined micro-mechanical objects that correspond to saddle points in the glass potential energy landscape. In this context, it is important to better understand both the origin of dilatational plastic instabilities and their outcomes, where the latter constitute the actual contribution to plastic deformation. Preliminary results (not shown here) indicate that the core properties of unstable plastic modes are correlated with the plastic strain accumulated as the glass reaches another potential energy minimum. Future work should systematically explore these correlations. In the context of the origin of plastic instabilities, the relations between quasilocalised, nonphononic modes in glasses (and other structural indicators~\cite{richard2020predicting}) to plastic instabilities under hydrostatic tension should be explored. In addition, the emergence of micro-cavitation, i.e., of particle-scale cavities, should be studied.

Second, at larger scales, future work should address the collective spatiotemporal organization of plastic deformation in glasses in the presence of hydrostatic driving forces, also going beyond the AQS limit (i.e., including finite temperatures and strain-rates). These should include the softening of the elastic moduli on the way to cavitation, demonstrated above, as well as large-scale cavity formation and subsequent dynamics. The coupling and competition between shear and dilatational plasticity should be addressed, including the relations and interplay between shear-banding and cavitation (e.g.,~\cite{richard2023bridging}). The emerging insight about dilatational plasticity should be eventually incorporated into coarse-grained elasto-plastic models, which will enable to address the fracture toughness of glasses (i.e., irreversible deformation and damage near the edges of crack defects~\cite{Rycroft_2012,Vasoya_2016}) and phenomena such as ductile-to-brittle transitions~\cite{ketkaew2018mechanical,richard2021brittle,richard2023bridging}.

\acknowledgements

A.M.~acknowledges support from the Minerva center on ``Aging, from physical materials to human tissues''. E.B.~acknowledges support from the Ben May Center for Chemical Theory and Computation and the Harold Perlman Family.

\appendix

\section{Methodology}
\label{sec:methods}

In this Appendix, we describe in more detail the methodology used in this work and the motivation behind it. Additional technical details are offered in~\cite{SM}. We employed in this work computer models to address elementary plasticity processes in glasses. Large-scale computer simulations played crucial roles in various recent developments in glass physics (see, for example,~\cite{falk_review,LB_swap_prx}). There are multiple reasons for this situation; first, computer glasses provide access to particle-level spatial scales that are essential for understanding glass plasticity, yet they are inaccessible through cutting-edge, real-time experimental techniques.

Second, computer simulations allow to control interatomic interactions in a way that goes well beyond current laboratory techniques. Particularly relevant for understanding dilatational plasticity is the ability to control the strength of the cohesive/attractive part of the interatomic interaction, as noted above. Specifically, we employed potential energy functions $U(\xv)$, where $\xv$ are the particle coordinates, composed of central force interatomic interactions of the Lennard-Jones type (see detailed formulation in~\cite{SM}) in which the cohesive/attractive strength is continuously adjustable, through the dimensionless parameter $r_{\rm c}$~\cite{karmakar2011effect,sticky_spheres_part_1,sticky_spheres_part2,SM} introduced above. The effect of varying $r_{\rm c}$ on various mesoscopic~\cite{sticky_spheres_part_1,sticky_spheres_part2} and macroscopic~\cite{richard2021brittle} quantities has been recently studied. Among these, we highlighted its effect on the fracture toughness of glasses, where reducing $r_{\rm c}$ can lead to a ductile-to-brittle transition~\cite{richard2021brittle}.

Third, computer glass simulations can be performed under athermal quasi-static (AQS) conditions~\cite{Malandro_Lacks,maloney2004subextensive,barriers_lacks_maloney,lemaitre2006_avalanches}, corresponding to the zero temperature and strain-rate limits. The AQS protocol is a powerful tool for studying fundamental aspects of the micro-mechanics, geometry and statistical-mechanical properties of glassy deformation. Its main merit is that it allows to exhaustively and unambiguously identify discrete plastic processes along the entire deformation path, as done in this work. Finally, computer glasses offer great advantages in implementing the deformation protocols described in Fig.~\ref{fig:shear_vs_dilation}. In the context of simple shear deformation, cf.~Fig.~\ref{fig:shear_vs_dilation}a, employing periodic boundary conditions allows to eliminate surface effects and hence reach very large strains without failure. Moreover, the hydrostatic tension test, cf.~Fig.~\ref{fig:shear_vs_dilation}b, allows to represent a mesoscopic portion of a macroscopic glass that experiences predominantly hydrostatic forces exerted by the surrounding material. We also note in passing that the application of isotropic dilation, which is readily accessible on the computer, is not easy to realize experimentally on the global scale (e.g., compared to the uniaxial tension test~\cite{richard2023bridging}), but is feasible (e.g.,~\cite{dorfmann2003stress}).

Glass samples in 3D, each with a fixed number of particles $N$ and initial volume $V_0$, were generated by rapidly quenching high-temperature, equilibrium binary-mixture liquids into zero temperature inherent states, as detailed in~\cite{SM}. While the non-equilibrium thermal history (or more generally thermo-mechanical history) of a glass has profound implications for its glassy state of disorder (e.g.,~\cite{gonzalez2023variability}), and correspondingly for its material properties --- plastic deformability in particular ---, we did not vary it in this work. It is important to note that our glasses feature vanishingly small initial pressure (as evident in Fig.~\ref{fig:shear_vs_dilation}b), which for the fixed quenching protocol is achieved by tuning $V_0$ (at a given $N$). This is important for revealing the intrinsic dilatational plasticity response of glasses.

Simple shear deformation (cf.~Fig.~\ref{fig:shear_vs_dilation}a) in a given direction is controlled by a shear strain $\gamma$. The latter is obtained through the accumulation of AQS strain increments $d\gamma$~\cite{SM}, for which the deformation gradient tensor ${\bm F}_{\rm s}\=\bm{\mathcal{I}}+d\gamma\,\hat{\bm x} \otimes\hat{\bm y}$
is applied (here $\bm{\mathcal{I}}$ is the identity tensor in 3D, $\hat{\bm x}$ and $\hat{\bm y}$ are Cartesian unit vectors, and $\otimes$ is a dyadic/outer product). The stress tensor, and in particular the simple shear stress component $\sigma(\gamma)$, as well as the shear $\mu(\gamma)$ and bulk $K(\gamma)$ moduli, were extracted~\cite{SM}. 

Hydrostatic tension (pure dilation, cf.~Fig.~\ref{fig:shear_vs_dilation}b) is controlled by a dilatational (volumetric) strain $\epsilon$. The latter is obtained through the accumulation of AQS strain increments $d\epsilon$~\cite{SM}, for which the deformation gradient tensor ${\bm F}_{\rm d}\=(1+d\epsilon)\,\bm{\mathcal{I}}$ is applied. The stress tensor, and in particular the hydrostatic tension (negative pressure) $-p(\epsilon)$, as well as the shear $\mu(\epsilon)$ and bulk $K(\epsilon)$ moduli, are extracted. 

The potential energy $U(\xv)$ of the glass is minimized during AQS deformation, controlled either by $\gamma$ or $\epsilon$. As shown above, and consistently with~\cite{dattani2022universal}, at least part of the plastic deformation in glassy materials --- independently of whether obtained under simple shear or hydrostatic tension (or more complex stress states) --- occurs through the accumulation of discrete plastic rearrangements (instabilities) that correspond to a zero crossing of an eigenvalue of the Hessian ${\calBold M}\!\equiv\!\frac{\partial^2U}{\partial\xv\partial\xv}$ (saddle-node bifurcation~\cite{Malandro_Lacks,barriers_lacks_maloney}).

The corresponding particle-level eigenfunctions/eigenmodes ${\bm u}({\bm r})$ (where ${\bm r}$ is a position vector relative to the center of the eigenmode~\cite{SM}), is accurately identified and extracted under AQS conditions. The plastic rearrangements (modes) ${\bm u}({\bm r})$ feature a highly-nonlinear, disordered core of linear size $a$ (i.e., of volume ${\cal V}\!\propto\!a^3$) and a power-law decay $|{\bm r}|^{-2}$ in the far field (e.g.,~\cite{lemaitre2006_avalanches}), $|{\bm r}|\!\gg\!a$, associated with a linear elastic continuum behavior~\footnote{We note that in certain limiting cases, e.g., related to the jamming point, the core can be extended and feature a power-law decay of its own, see~\cite{lerner2023anomalous}}. The irreversible deformation occurs inside the nonlinear core, whose averaged effect is quantified through an eigenstrain tensor $\bm{\mathcal{E}}^*$ in the framework of Eshelby's inclusions formalism~\cite{eshelby1957determination,eshelby1959elastic}. 

A recently developed method~\cite{moriel2020extracting}, used in this work, allows to extract ${\cal V}\,\bm{\mathcal{E}}^*$. Earlier approaches, discussed in the works of~\cite{ashwin2013yield,albaret2016mapping,boioli2017shear,nicolas2018orientation} (some of which are discussed above), also invoked Eshelby's inclusions in similar contexts. We note that we analyzed the unstable modes very close to the saddle-node bifurcation, which are well defined mathematical objects, and not the outcome of instability obtained once a new energy minimum is reached. The relation between the two was briefly discussed above. Note also that we considered simple shear and hydrostatic tension as end members of a continuum of stress states, which can be studied as well.

\clearpage

\onecolumngrid

\begin{center}

\textbf{\large Supplemental material for: \\``Elementary processes in dilatational plasticity of glasses''}

\end{center}




\setcounter{equation}{0}

\setcounter{figure}{0}

\setcounter{section}{0}

\setcounter{subsection}{0}

\setcounter{table}{0}

\setcounter{page}{1}

\makeatletter

\renewcommand{\theequation}{S\arabic{equation}}

\renewcommand{\thefigure}{S\arabic{figure}}

\renewcommand{\thesubsection}{S-\Roman{subsection}}

\renewcommand*{\thepage}{S\arabic{page}}





\twocolumngrid

\title{Supplemental material for: \\``Elementary processes in dilatational plasticity of glasses''}
\author{Avraham Moriel$^{1}$}
\author{David Richard$^{2}$}
\author{Edan Lerner$^{3}$}
\author{Eran Bouchbinder$^{1}$}
\affiliation{$^{1}$Chemical and Biological Physics Department, Weizmann Institute of Science, Rehovot 7610001, Israel\\
$^{2}$Universit\'{e} Grenoble Alpes, CNRS, LIPhy, 38000 Grenoble, France\\
$^{3}$Institute for Theoretical Physics, University of Amsterdam, Science Park 904, Amsterdam, Netherlands}
\maketitle

In this Supplemental materials file, we provide additional technical details regarding the quantities used and computations reported on in the manuscript. In particular, we provide details regarding the interaction potentials and the glass preparation protocol. We also provide details regarding the plastic instabilities detection algorithm, and contour integrals method for extracting the core properties of plastic instabilities. Finally, we present some additional results that support statements made in the manuscript.

\section{Interaction potentials and preparation protocol}

To study the effects of variation in interaction potential, we used binary mixtures of small and large particles, interacting via two variations of the sticky-sphere potential, introduced and studied in~\cite{karmakar2011effect, Massimo_supercooled_PRL,sticky_spheres_part_1, sticky_spheres_part2}. The pairwise potential takes the form
\begin{widetext}
\begin{equation}\label{eq:ss}
    \phi_{ij}=\begin{cases}
4\varepsilon\left[\left(\frac{\lambda_{ij}}{r_{ij}}\right)^{12}-\left(\frac{\lambda_{ij}}{r_{ij}}\right)^{6}\right]\ , & \frac{r_{ij}}{\lambda_{ij}}<x_{{\rm min}}\ ,\\
\varepsilon\left[a\left(\frac{\lambda_{ij}}{r_{ij}}\right)^{12}-b\left(\frac{\lambda_{ij}}{r_{ij}}\right)^{6}+\sum_{\ell=0}^{3}c_{2\ell}\left(\frac{r_{ij}}{\lambda_{ij}}\right)^{2\ell}\right]\ , & x_{{\rm min}}\le\frac{r_{ij}}{\lambda_{ij}}<x_{c}\\
0\ , & \frac{r_{ij}}{\lambda_{ij}}\ge0
\end{cases}
\end{equation}
\end{widetext}
with $\varepsilon$ being a microscopic energy scale, $r_{ij}\!\equiv\!\left| \bm{r}_i - \bm{r}_j\right|$ is the distance between particles $i$ and $j$, $x_{\rm{min}}\!=\!2^{1/6}$ is the dimensionless location of the minimum of the potential, $x_c$ is a modified cutoff, and $\lambda_{ij}$ is an interaction length, expressed in terms of the ``small-small'' interaction length $\lambda_{\rm{small}}^{\rm{small}}$, $\lambda_{\rm{small}}^{\rm{large}}\!=\!1.18\lambda_{\rm{small}}^{\rm{small}}$, and $\lambda_{\rm{small}}^{\rm{large}}\!=\!1.4\lambda_{\rm{small}}^{\rm{small}}$. The parameters $a,b,\{c_{2\ell}\}$ detailed in Table~\ref{ta:rcs} are set such that the attractive and repulsive parts of $\phi_{ij}$, as well as its first two derivatives are continuous at $x_{\rm{min}}$ and $x_c$. We refer to the two sticky-sphere variations used in our work by defining $r_{\rm c}\!\equiv\!x_c/x_{\rm{min}}$. When plotting the potentials in the inset of Fig.~2a in the manuscript, we have used $r\!=\!r_{ij} / \lambda_{ij}$, and the normalized $\tilde{r}\!=\!r/x_{\rm{min}}$ such that the minimum occur at $\tilde{r}\!=\!1$~\cite{sticky_spheres_part_1, sticky_spheres_part2}.
\begin{table}
\begin{tabularx}{0.4\textwidth}{|X X|}
\hline
\multicolumn{2}{|c|}{$r_{\rm c} = 1.2$} \\
\hline\hline
$a$ & -106.991613526652 \\
$b$ & -304.918469059567 \\
$c_0$ & -939.388037994211 \\
$c_2$ & 1190.70962256002 \\
$c_4$ & -541.3001315875512 \\
$c_6$ & 85.86849369147127 \\
\hline\hline
\multicolumn{2}{|c|}{$r_{\rm c} = 1.5$} \\
\hline\hline
$a$ & 1.1582440286928275\\
$b$ & -2.2619482444770567\\
$c_0$ & -12.414700446492716\\
$c_2$ & 12.584354590303674\\
$c_4$ & -4.320508006050397\\
$c_6$ & 0.49862551162881885\\
\hline
\end{tabularx}
\caption{Parameters of the sticky-sphere potential, cf.~Eq.~\eqref{eq:ss}, with $r_{\rm c}\!=\!1.2$, and $r_{\rm c}\!=\!1.5$, respectively.}
\label{ta:rcs}
\end{table}

We have initialized 200 samples of 50:50 binary mixtures with $N\!=\!128$K and equilibrated them at a high parent temperature $T_{\rm p}\!=\!4.0$ employing the Berendsen thermostat~\cite{berendsen}. Then, we quenched the samples instantaneously to zero temperature, followed by overdamped dynamics. To obtain the illustrative stress-strain curves of Fig.~1 in the manuscript, we used 108 independent systems of $N\!=\!10$K particles and performed athermal quasi-static deformations to large strains, i.e, beyond the yielding transition.

Most studies in the literature focus on simple shear deformations, where little attention is given to the initial pressure resulting from the glass preparation protocol. However, as we are interested in applying both simple shear and hydrostatic tension to the same glass realizations, we aim at ensuring that the initial pressure is small, preferably vanishingly small compared with the as-cast bulk modulus $K$. This is important for revealing the intrinsic dilatational plasticity response of glasses, which is commonly probed in the laboratory at ambient (hence, small) pressure (obviously varying the initial pressure may affect the subsequent response to hydrostatic tension). This is achieved by tuning the initial number density for the different $r_{\rm c}$ values (for our fixed quenching protocol). Using the densities stated in Table~\ref{ta:quench} before the quenching procedure, we attain the stated pressures at the end of the quench, without any additional processing.

Details regarding the initial densities, and resulting pressures appear in Table~\ref{ta:quench}. In particular, we have $p(\epsilon\!=\!0)/K(\epsilon\!=\!0)\!\simeq\!{\cal O}(10^{-3})$, where $K(\epsilon\!=\!0)$ is the initial bulk modulus (to be discussed in Sec.~\ref{sec:moduli_integrals}), for both $r_{\rm c}$'s and across samples.

\begin{table}
\begin{tabular}{|c||c|c|c|}
\hline
 $r_{\rm c}$ & $\rho$ &  $p(\epsilon\!=\!0) / K(\epsilon\!=\!0)$ & $\nu(\epsilon\!=\!0)$ \\ \hline
 \hline
 $r_{\rm c} = 1.2$ &  $0.528$ & $ (2.8 \pm 0.2)\cdot10^{-3} $ & $0.28$ \\ \hline
 $r_{\rm c} = 1.5$ & $0.536$ &  $(4.0 \pm 0.12)\cdot 10^{-3}$ & $0.38$\\ \hline
\end{tabular}
\caption{Initial density and resulting pressure (normalized by the bulk modulus $K$) and Poisson's ratio $\nu$ in the quenched samples for the two $r_{\rm c}$ values.}
\label{ta:quench}
\end{table}

\section{Loading and event detection}
\label{supse:detect}
\begin{figure}[b!]
  \centering
  \includegraphics[width=0.48\textwidth]{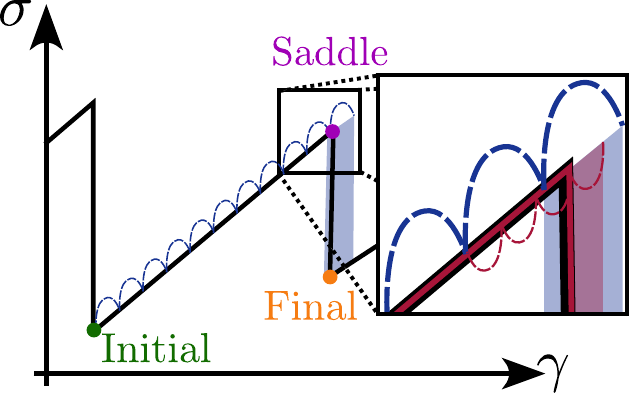}
  \caption{An illustration of the first two stages of the plastic instability detection algorithm. In the first stage, the algorithm starts from the initial state (green) and imposes athermal quasi-static deformation increments (blue dashed upper parabolas). It results in a coarse-level estimate of the instability (blue-shaded region). In the second stage, we adaptively decrease the strain increment (inset, red dashed lower parabolas), resulting in a refined estimate of the instability (red-shaded region). The adaptive refinement is applied iteratively until an accurate estimate of the instability strain is obtained, as described in detail in the text.}
  \label{fig:algorithm}
\end{figure}

As explained in the manuscript, we apply to our glass samples athermal quasi-static (AQS) deformation, either in simple shear or in hydrostatic tension. Under hydrostatic tension, periodic boundary conditions in all space directions are imposed. Under simple shear, we employ the Lees-Edwards boundary conditions in the shearing direction~\cite{allen2017computer}. Below, we specify how we apply deformations and detect instabilities under simple shear deformations, and will use $\gamma$ to denote the strain. We use the same procedure for loading and instability detection under hydrostatic tension (unless stated otherwise), such that $\gamma$ and $\epsilon$ are interchangeable.

The AQS deformation is performed conventionally, where at each step we apply a global affine transformation to the particle positions ${\bm x}$, quantified through a deformation gradient tensor ${\bm F}$, followed by energy minimization that results in non-affine displacements. As explained in the manuscript, we either use ${\bm F}_{\rm s}\=\bm{\mathcal{I}}+d\gamma\,\hat{\bm x} \otimes\hat{\bm y}$ for simple shear or ${\bm F}_{\rm d}\=(1+d\epsilon)\,\bm{\mathcal{I}}$ for hydrostatic tension. Here $\bm{\mathcal{I}}$ is the identity tensor in 3D, $\hat{\bm x}$ and $\hat{\bm y}$ are Cartesian unit vectors, and $\otimes$ is a dyadic/outer product. As stated above, we hereafter use $d\gamma$ to symbolically denote strain increments either under simple shear or hydrostatic tension.

Plastic instabilities under AQS conditions, which are the main micro-mechanical objects discussed and analyzed in the manuscript, correspond to vanishing eigenvalues of the glass Hessian $\bm{\mathcal{M}}\!\equiv\!\frac{\displaystyle \partial^2 U}{\displaystyle \partial\bm{x}\,\partial\bm{x}}$ evaluated at $\bm{x}(\gamma)$, i.e.~$\bm{\mathcal{M}}$ is a function of the imposed strain $\gamma$. Consequently, one's goal is to calculate $\bm{\mathcal{M}}(\gamma)$ and its lowest eigenvalue (i.e., diagonalize the Hessian) along the entire deformation path $\bm{x}(\gamma)$. This poses several serious technical challenges; first, it is computationally expensive to perform these computations along the entire deformation path. Second, it is crucial to very accurately detect the strain at the saddle point $\gamma^{\rm{s}}$ at which plastic instabilities take place. In particular, an a priori choice of strain increments $d\gamma$ would be most likely insufficient, i.e., too coarse.

The procedure we invoked to meet this challenge is illustrated in Fig.~\ref{fig:algorithm}. We highlight therein 3 states, an initial state (which either corresponds to $\gamma\=0$ or to a final state resulting from a previous instability), a saddle (that we aim at accurately detecting) and a final state (emerging as the outcome of the instability). The crucial point is the selection of the strain increments $d\gamma$, as mentioned above, which is obtained using 3 stages (the first two stages were already used in~\cite{MW_theta_and_omega}):

\noindent\, $\bullet$ In the first stage, the coarsest one, we initially set $d\gamma\=10^{-4}$,  marked by the blue dashed upper parabolas in Fig.~\ref{fig:algorithm}, and identify pairs of consecutive states $\bm{x}(\gamma)$ and $\bm{x}(\gamma+d\gamma)$. We then compute the non-affine displacement between each pair as $\bm{\Delta}(\gamma)\!\equiv\!\mathcal{T}_{\rm naf}\,[\bm{x}(\gamma+d\gamma)-\bm{x}(\gamma)] / \tilde{f}\left(\gamma\right)$. Here, the operator $\mathcal{T}_{\rm naf}$ removes the affine displacements and accounts for boundary effects (e.g., particles crossing periodic boundaries), and $\tilde{f}\left(\gamma\right)$ is the particle-averaged force induced by the affine transformation. The largest component of $\bm{\Delta}(\gamma)$ is denoted as $\Delta^{\text{max}}(\gamma)$. We repeat this process for each pair of states and consider the ratio $\Delta^{\mbox{\tiny max}}(\gamma + d\gamma)/\Delta^{\text{max}}(\gamma)$. Once the latter surpasses a threshold value $\Theta_0\!=\!2$, i.e., once the non-affine displacements significantly increase with strain, we deduce that a plastic instability took place. This is illustrated by the last blue upper parabola in Fig.~\ref{fig:algorithm}, implying that --- on this coarse level --- the instability is estimated to occur somewhere in between, represented by the blue-shaded region therein.

\noindent\, $\bullet$ In the second stage, which aims to refine the identification of the instability strain within the blue-shaded region, we apply an adaptive and iterative refinement process, as schematically illustrated in the inset of Fig.~\ref{fig:algorithm} by the red dashed lower parabolas. We consider a state obtained a few coarse strain steps earlier (before the coarse-level identification of the instability). We then redefine our threshold to be $\Theta\!\equiv\!\Delta(\gamma + d\gamma)^{\text{max}}/\Delta(\gamma)^{\text{max}}\!>\!\Theta_0$, halve the strain increment and repeat the deformation procedure of the first stage at this finer level.
We repeat this process of halving the strain increment and increasing the adaptive threshold ratio $\Theta$ as long as the latter is surpassed in each step (indicating that an instability is approached), until the strain increment drops below $10^{-6}$. The resulting state corresponds to our refined estimate of the saddle strain $\tilde{\gamma}^{\rm{s}}$, represented by the red vertical line in the inset of Fig.~\ref{fig:algorithm} and the associated red-shaded region. We denote the corresponding state as ${\bm x}^{\rm{s}}$. In addition, we push the system over the instability, which results in the final state (outcome of instability) marked in Fig.~\ref{fig:algorithm}. We denote the corresponding state as ${\bm x}^{\rm{fin}}$.

\noindent\, $\bullet$ In the third stage, the most refined one, we first solve for the lowest eigenvalue of the eigenvalue problem $\bm{\mathcal{M}}(\tilde{\gamma}^{\rm{s}}) \bm{\psi}\= \omega^2 \bm{\psi}$, where $\bm{\psi}$ is the eigenmode associated with the minimal frequency $\omega$. $\bm{\psi}$, being a harmonic eigenmode, may suffer from hybridization with phononic plane waves, as extensively discussed in the literature~\cite{Wijtmans2017,Kapteijns2020,QLE_jcp_2021}. We bypass this potential difficulty by employing a recently developed \emph{cubic} nonlinear-modes framework that cleanly and effectively disentangles plastic instabilities from the phononic background, as detailed in~\cite{micromechanics2016,SciPost2016}. We use $\bm{\psi}$ as an input to the nonlinear mode calculation, resulting in a displacement field ${\bm u}({\bm r})$, which is the basic quantity used in the manuscript to extract the properties of plastic instabilities. An added value of the nonlinear-modes framework, as discussed in~\cite{micromechanics2016,SciPost2016}, is that it converges more accurately to the instability, and hence effectively provides us with an even better estimate of ${\gamma}^{\rm{s}}$ (compared to $\tilde{\gamma}^{\rm{s}}$), which was our overall goal.

In addition, we are interested in the sign of ${\bm u}({\bm r})$, which allows to distinguish between isotropic core expansion/dilation, $\epsilon_{\mbox{\scriptsize dil}}^*\!>\!0$, and isotropic core contraction, $\epsilon_{\mbox{\scriptsize dil}}^*\!<\!0$ (also rendering the plastic eigenstrain triaxiality $R_{\rm t}$ in Eq.~(1) and Fig.~2 in the manuscript a \emph{signed} quantity). Therefore, we fix the overall sign of ${\bm u}({\bm r})$ such that ${\bm u}\!\cdot\! (\bm{x}^{\rm{fin}}-\bm{x}^{\rm{s}})\!>\!0$.

The 3-stage procedure detailed above allows us to obtain a set of initial and unstable (saddle) states, where the latter are analyzed in the manuscript. Finally, the shear strain $\gamma$ (and dilatational strain $\epsilon$), as appearing on the $x$ axis in Figs.~1 and~5 in the manuscript, correspond to the accumulated effect of the discrete strain steps involved in the procedure (the coarse and the adaptively refined ones).
\begin{figure}[b!]
  \centering
  \includegraphics[width=0.48\textwidth]{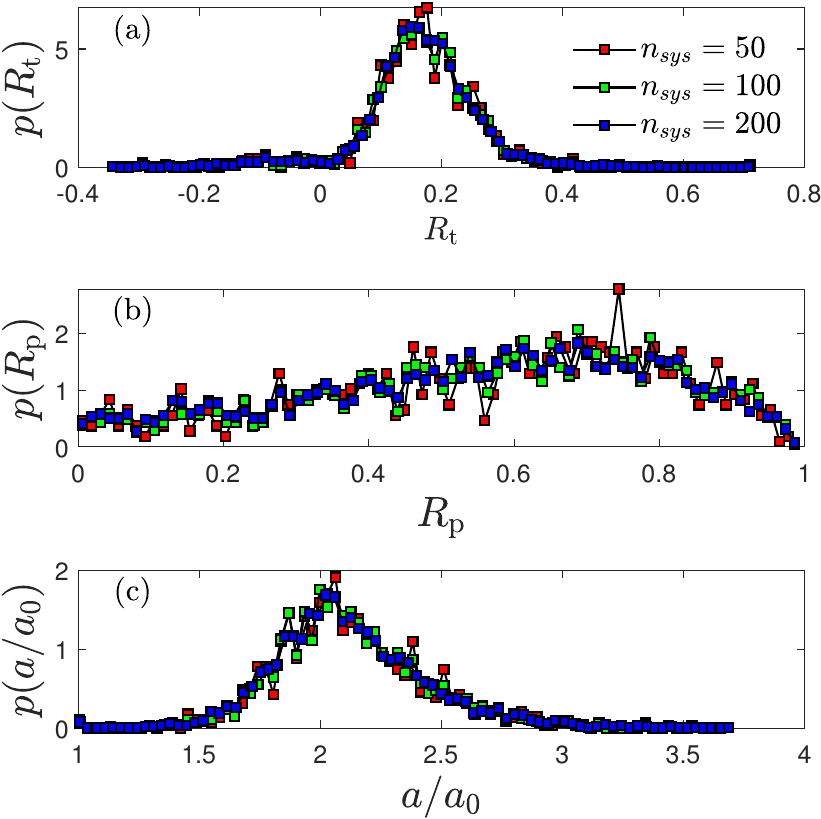}
  \caption{A demonstration of statistical convergence of $r_{\rm c}\!=\!1.2$ under hydrostatic tension. The distribution of (a) $R_{\rm t}$, (b) $R_{\rm p}$ and (c) $a/a_0$ for 50 systems (analysis of $876$ modes) in red, for 100 systems ($1673$ modes) in green, and 200 systems ($3345$ modes) in blue. All quantities appear to converge even for this relatively small ensemble of modes.}
  \label{fig:convergence}
\end{figure}

\section{Extracted quantities: elastic moduli and core properties through contour integrals}
\label{sec:moduli_integrals}

For each initial state, we compute the shear and bulk moduli according to~\cite{lutsko,sticky_spheres_part_1}
\begin{subequations}
    \begin{eqnarray}
    \label{eq:mu}
  \hspace{-1cm}  \mu(\bm{\epsilon}_{\rm{fin}}) =& \frac{\displaystyle \frac{\partial^2 U}{\partial \gamma^2} - \frac{\partial^2 U}{\partial \gamma \partial \bm{x}}\cdot \bm{\mathcal{M}}\cdot \frac{\partial^2 U}{\partial \bm{x} \partial \gamma}}{\displaystyle V} \ , \\
   \label{eq:K}
  \hspace{-1cm}   K(\bm{\epsilon}_{\rm{fin}}) =& \frac{\displaystyle\frac{\partial^2 U}{\partial\epsilon^2} - \dbar \frac{\partial U}{\partial \epsilon} - \frac{\partial^2 U}{\partial \epsilon \partial \bm{x}} \cdot \bm{\mathcal{M}}\cdot \frac{\partial^2 U}{\partial \bm{x} \partial \epsilon }}  {\displaystyle V\,\dbar^2} \ .
    \end{eqnarray}
\end{subequations}
Here, derivatives are taken with respect to $\gamma$, in Eq.~\eqref{eq:mu}, and to $\epsilon$, in Eq.~\eqref{eq:K}. Both $\gamma$ and $\epsilon$ have the same geometric meaning as in the corresponding global affine transformations used in the actual deformation, though here they are used as putative motions invoked to evaluate linear response coefficients, independently of whether the actual deformation is simple shear or hydrostatic tension. $\dbar$ is the spatial dimension, which in our work is set to $\dbar\!=\!3$.

Note that the elastic moduli in Eqs.~\eqref{eq:mu}-\eqref{eq:K} include an overall factor $1/V$. Under simple shear, we have $V(\gamma)\=V_0$, independently of the shear strain $\gamma$. However, under hydrostatic tension, $V(\epsilon)\!\ge\!V_0$ is an increasing function of the dilatational strain $\epsilon$. In the inset of Fig.~5b in the manuscript, we present the reduced moduli under hydrostatic tension, defined as $\tilde\mu(\epsilon)\!\equiv\!\mu(\epsilon)V(\epsilon)/V_0$ and $\tilde K(\epsilon)\!\equiv\!K(\epsilon)V(\epsilon)/V_0$.

The elastic moduli are subsequently employed in the contour integrals method developed in~\cite{moriel2020extracting}, used to extract the the core properties ${\cal V}\,\bm{\mathcal{E}}^*$ of each unstable plastic mode $\bm{u}(\bm{r})$. The contour integrals is presented in great detail in~\cite{moriel2020extracting} and we apply it in the present work as is. For completeness, we briefly note that ${\cal V}\,\bm{\mathcal{E}}^*$ are obtained via a specifically designed  projection of the mode $\bm{u}(\bm{r})$ onto the spherical harmonics~\cite{moriel2020extracting}
\begin{subequations}
    \begin{eqnarray}
      \hspace{-1cm}  I_0(r)\!\!&\equiv&\!\! 2\sqrt{\pi} \left(\frac{\lambda + 2\mu}{3\lambda + 2\mu}\right) \int_S \bm{u}(\bm{r}) \cdot \bm{r}\,Y_0^0(\Omega)\,r\,d\Omega \ , \\
      \hspace{-1cm}  I_2^{(m)} \!\!&\equiv&\!\! 2\sqrt{5\pi} \left(\frac{\lambda + 2\mu}{3\lambda + 5 \mu}\right) \int_S \bm{u}(\bm{r}) \cdot \bm{r}\,Y_2^m(\Omega)\,r\,d\Omega \ ,
    \end{eqnarray}
\end{subequations}
where $\lambda\!=\! K - 2\mu/3$, the surface integral $S$ is performed on a sphere of radius $r$, $Y_0^0(\Omega)$ and $Y_2^m(\Omega)$ are the real orthogonal spherical harmonics of the zeroth and second degree and orders zero and $m\!=-2,-1,0,1,2$~\cite{Blanco1997,chisholm1976} and $\Omega$ is the solid angle. To obtain ${\cal V}\,\bm{\mathcal{E}}^*$ we then construct the matrix $\bar{\bm{I}}_2(r)$ as detailed in~\cite{moriel2020extracting}, and extract the core properties at a distance $r$ where the largest eigenvalue of $\bar{\bm{I}}_2(r)$ in absolute value attains its maximum.

\begin{figure}[]
  \centering
  \includegraphics[width=0.48\textwidth]{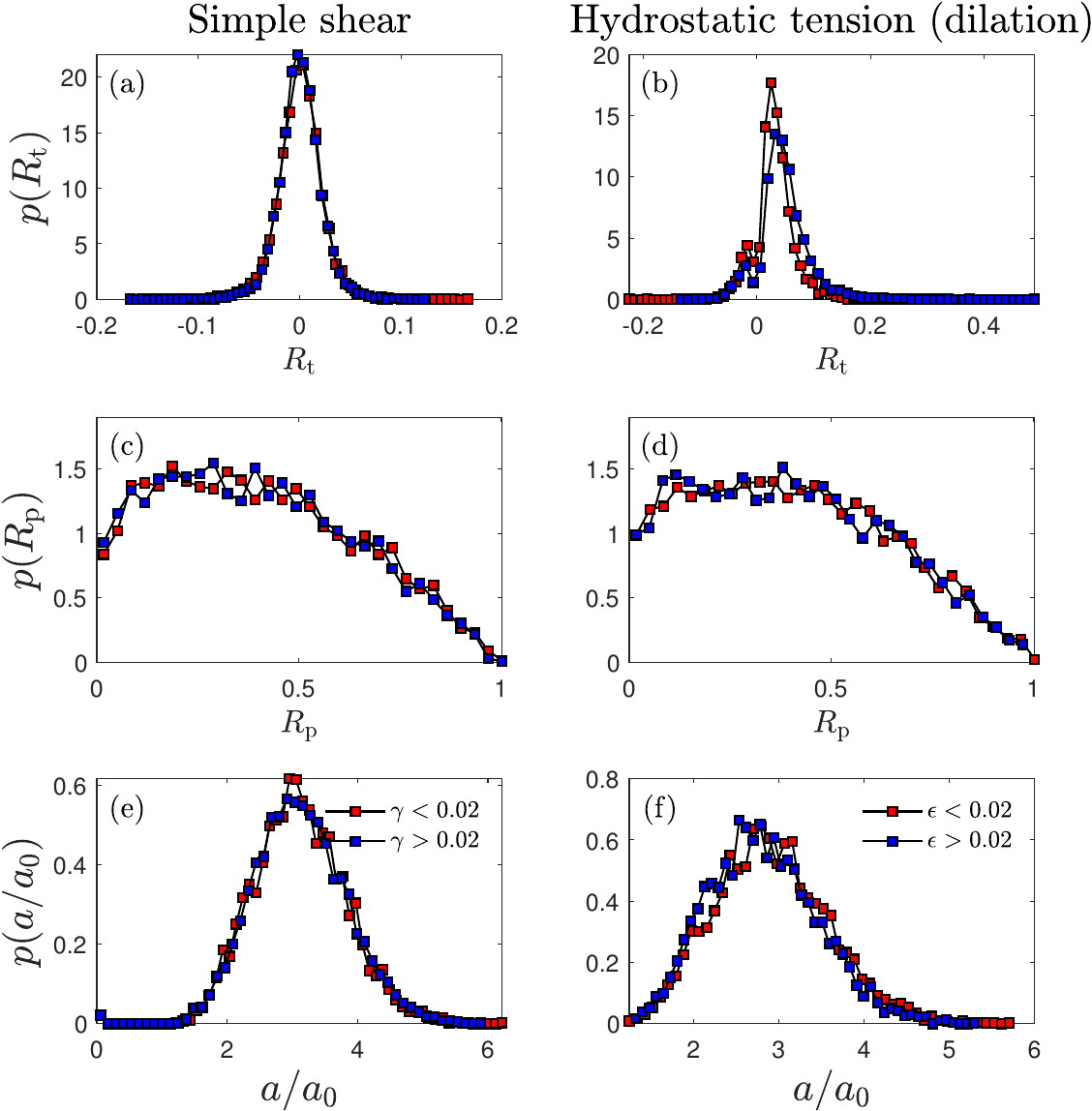}
  \caption{Distributions of (a-b) $R_{\rm t}$,  (c-d) $R_{\rm p}$, and (e-f) $a/a_0$, for $r_{\rm c}\!=\!1.5$ at early and late stages of loading under simple shear (left column) and hydrostatic tension (right column). For simple shear, we used 4039 modes for the early strain regime and 5728 for the developed strain regime, and for hydrostatic tension, we used 5007 for the early strain regime and 4992 for the developed strain regime. Overall, we do not observe any significant change in distributions, except for a mild change in the histogram in (b) for $R_{\rm t}$ under dilation, which is somewhat anticipated.}
  \label{fig:strain_1.5}
\end{figure}
\begin{figure}[h!]
  \centering
  \includegraphics[width=0.48\textwidth]{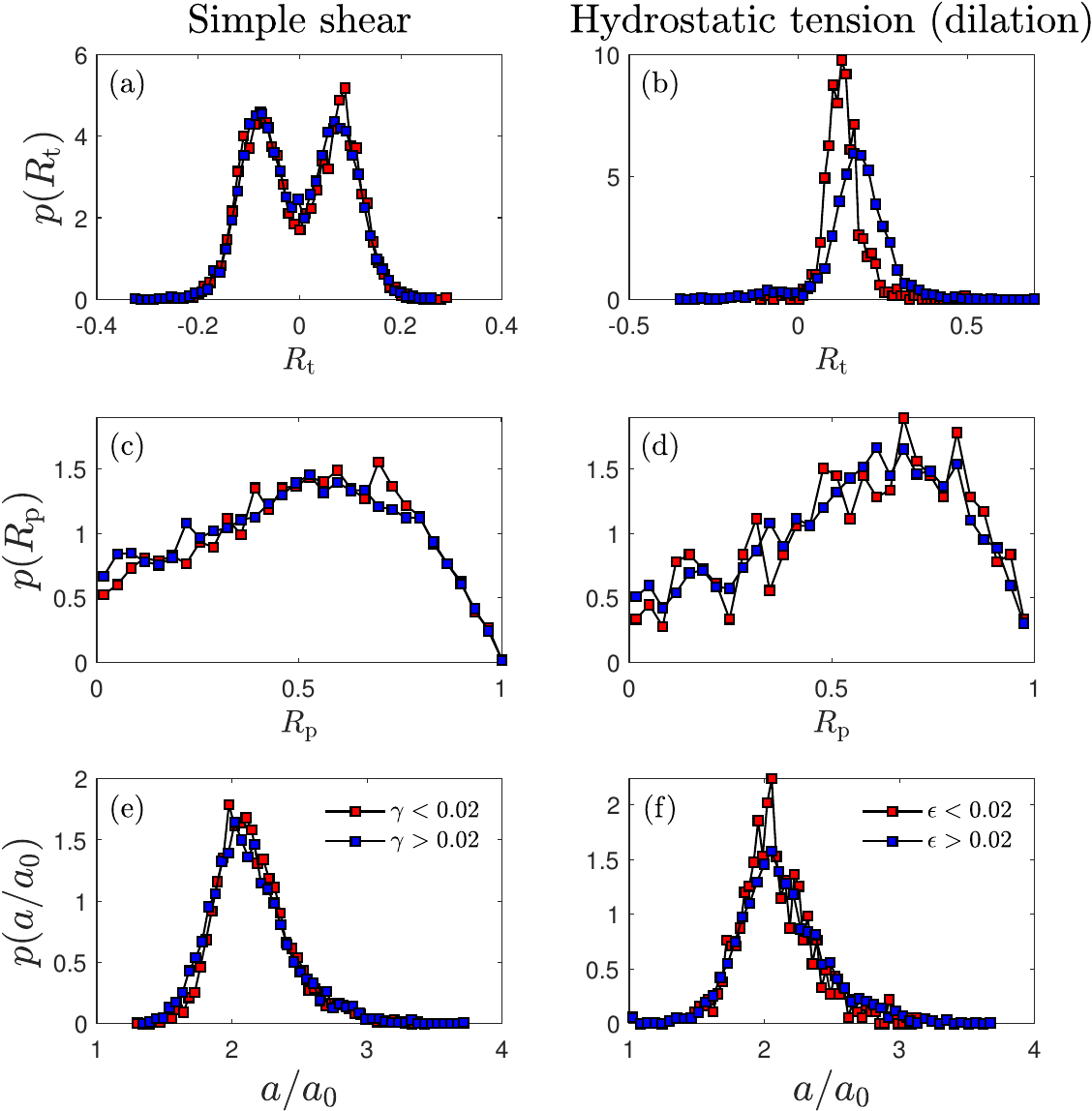}
  \caption{Distributions of (a-b) $R_{\rm t}$,  (c-d) $R_{\rm p}$, and (e-f) $a/a_0$, for $r_{\rm c}\!=\!1.2$ at early and late stages of loading under simple shear (left column) and hydrostatic tension (right column). For simple shear, we used 4544 modes for the early strain regime and 5456 for the developed strain regime, and for hydrostatic tension, we used 544 for the early strain regime and 2801 for the developed strain regime. Overall, we do not observe any significant change in distributions, except for a mild change in the histogram in (b) for $R_{\rm t}$ under dilation, which is somewhat anticipated.}
  \label{fig:strain_1.2}
\end{figure}

\section{Statistical convergence and weak strain dependence}

As explained in the manuscript, there is some variability in the number of plastic events used for the statistical analysis between different $r_{\rm c}$ values and loading symmetries. The smallest data set corresponds to $r_{\rm c}\!=\!1.2$ under hydrostatic tension, in which we have collected 3345 events for analysis. In this context, it is important to verify the statistical convergence of the statistical distributions by considering smaller subsets of each ensemble, as demonstrated in Fig.~\ref{fig:convergence}.

We also examined the plastic eigenstrain triaxiality ratio $R_{\rm t}$, planarity ratio $R_{\rm p}$, and core size $a/a_0$ as the strain increased throughout the deformation path. Figure~\ref{fig:strain_1.5} shows the obtained distributions for $R_{\rm t}$, $R_{\rm p}$, and $a/a_0$ originating from analyzing modes at the initial and largest-strain phases of loading ($\gamma\!<\!0.02$ and $\gamma\!>\!0.02$ under simple shear, and $\epsilon\!<\!0.02$ and $\epsilon\!>\!0.02$ under hydrostatic tension, respectively) for $r_{\rm c}\!=\!1.5$. It is observed that most distributions are strain independent, except for a weak dependence of $p(R_{\rm t})$ under hydrostatic tension (panel (b)). Figure~\ref{fig:strain_1.2} shows the same distributions for $r_{\rm c}\!=\!1.2$. The results are similar to those for $r_{\rm c}\!=\!1.5$, indicating weak strain dependence, yet with a somewhat larger effect in $p(R_{\rm t})$ under hydrostatic tension (panel (b)). In general, we expect strain dependence to emerge under hydrostatic tension for larger deformation level, an issue to be addressed in future work.


\end{document}